\documentclass{aa}  

\usepackage{graphicx}
\usepackage[utf8]{inputenc}
\usepackage{hyperref}
\usepackage{ae,aecompl}

\newcommand{\kms}{\,km\,s$^{-1}$} 
\newcommand{\ms}{\,m\,s$^{-1}$} 
\newcommand{\Ppuls}{$P_{\mathrm{puls}}$} 
\newcommand{\Porb}{$P_{\mathrm{orb}}$} 

\bibpunct{(}{)}{;}{a}{}{,} 

\begin{document}

   \title{Probing Polaris' Puzzling Radial Velocity Signals}
   \subtitle{Pulsational (In-)Stability, Orbital Motion, and Bisector Variations}

   \author{R.~I. Anderson
          \inst{1,2}\fnmsep\thanks{ESO fellow;  \email{randerso@eso.org}}
          }

   \institute{European Southern Observatory, Karl-Schwarzschild-Str. 2, 85748 Garching b. M\"unchen, Germany
    \and
             D\'epartement d'Astronomie, Universit\'e de Gen\`eve, 51 Ch. des Maillettes, 1290 Sauverny, Switzerland                        }

   \date{Received 22 November 2018; revised 8 February 2019; accepted 18 February 2019}

  \abstract
   {
   We investigate temporally changing variability amplitudes and the multi-periodicity of the type-I Cepheid Polaris using 161 high-precision radial velocity (RV) and bisector inverse span (BIS) measurements based on optical spectra recorded using {\it Hermes} at the 1.2\,m Flemish Mercator telescope on La Palma, Canary Islands, Spain.
   Using an empirical template fitting method, we show that Polaris' RV amplitude has been stable to within $\sim 30$\ms\ between September 2011 and November 2018. 
   We apply the template fitting method to publicly accessible, homogeneous RV data sets from the literature and provide an updated solution of Polaris' eccentric $29.3$\,yr orbit. 
   While the inferred pulsation-induced RV amplitudes differ among individual data sets, we find no evidence for time-variable RV amplitudes in any of the separately considered, homogeneous data sets. Additionally, we find that increasing photometric amplitudes determined using {\it SMEI} photometry are likely spurious detections due to as yet ill-understood systematic effects of instrumental origin. Given this confusing situation, further analysis of high-quality homogeneous data sets with well-understood systematics is required to confidently establish whether Polaris' variability amplitude is subject to change over time. 
   We confirm periodic bisector variability periods of $3.97$\,d and $40.22$\,d using {\it Hermes} BIS measurements and identify a third signal at a period of $60.17$\,d. Although the $60.17$\,d signal dominates the BIS periodogram, we caution that this signal may not be independent of the $40.22$\,d signal. Finally, we show that the $40.22$\,d signal cannot be explained by stellar rotation.
   Further long-term, high-quality spectroscopic monitoring is required to unravel the complete set of Polaris' periodic signals, which has the potential to provide unprecedented insights into the evolution of Cepheid variables.
   }

   \keywords{Stars: Individual: Polaris = North Star = $\alpha$ UMi = HD\,8890 -- Stars: variables: Cepheids -- binaries: spectroscopic -- binaries: visual -- Stars: oscillations -- techniques: radial velocities
               }

   \maketitle
%

\section{Introduction \label{sec:introduction}}

The North Star, \object{Polaris}\footnote{Unless otherwise stated, Polaris shall refer to the Cepheid variable component Aa of the $\alpha$ UMi system using the nomenclature of \citet{2008AJ....136.1137E}.}, is a celebrity among type-I Cepheid variable stars (henceforth: Cepheids) thanks to its proximity, uncertain physical properties, and plentiful literature concerning its puzzling variability properties. 
Being the closest Cepheid to the Sun, one might expect Polaris to have a prototypical role in the understanding of Cepheids in general, which are both important stellar laboratories and accurate cosmic yardsticks. In particular the application of Cepheids as standard candles for determining the value of the Hubble-Lema\^itre constant with high accuracy \citep{2018ApJ...861..126R,2016ApJ...826...56R} provides a strong motivation for achieving a solid understanding of their astrophysical properties. Yet, Polaris continues to defy a detailed description, with several recent articles struggling to explain its mass, radius, age, and other properties \citep[cf.][]{2018ApJ...853...55B,2018A&A...611L...7A,2018ApJ...863..187E}. Most recently, \citet{2018MNRAS.481L.115U} have refueled previous discussions on the stability of Polaris' variability, which may be the key to explaining this uncomfortable situation.

Polaris' weak photometric variability was first noticed in the mid 19th century \citep{seidel1852untersuchungen} and was only fully confirmed in the early 20th century \citep{1911AN....189...89H,1912PA.....20..505K,1913AN....194..359P}. Early radial velocity (RV) observations using the the Mills spectrograph at Lick Observatory \citep{1898ApJ.....8..123C} revealed RV variability on two timescales, leading \citet{1899ApJ....10..180C} to conclude that Polaris ``is at least a triple system\footnote{the $4$\,d periodicity due to pulsation was interpreted as orbital motion at the time}''. Thirty years later \citet{1929PASP...41...56M} determined an orbital period, \Porb, of 29 years. 

\citet{1965ApJ...141.1415R} confirmed this orbit and noted irregularity in the periodicity of the RV variations and adopted different photometrically determined pulsation periods, \Ppuls, for different epochs of RV measurements. Indeed, the rate of period change is exceptionally high for Polaris \citep[$\dot{P} \sim 4.5\, \rm{s\,yr^{-1}}$]{2005PASP..117..207T}, which implies a first crossing the classical instability strip, i.e., that Polaris populates the Hertzsprung gap and has not yet undergone the first dredge-up event \citep{2016A&A...591A...8A}. \citet{2005PASP..117..207T} further noticed an abrupt change to \Ppuls\ between 1963 and 1966, which cannot be explained using the standard interpretation of secular evolution. 

\citet{1983ApJ...274..755A} reported first indications of a time-variable amplitude, suggesting that Polaris was about to exit the instability strip. Despite focusing on photometric observations, Arellano Ferro also made brief reference to a significant decrease in RV amplitude between the Lick velocities ($2K \sim 6$\kms) and newer RVs measured ($2K \sim 2$\kms) at the David Dunlap Observatory (DDO) \citep[cf. also][]{1984JRASC..78..173K}. This report sparked much interest in Polaris, leading to confirmations of a decreased RV amplitude and predictions that Polaris would cease pulsating in the mid 1990s \citep{1989AJ.....98.2249D,1993ApJ...416..820F}. However, Polaris' RV amplitude appeared to stabilize by late 1997 \citep{1998AJ....116..936K,2000AJ....120..979H} and even showed a slow but noticeable increase in between 2004 and 2007 \citep{2008AJ....135.2240L,2008ApJ...683..433B}. Most recently, \citet{2018MNRAS.481L.115U} reported a significantly higher RV amplitude around $3.8 - 2.8$\kms, which they interpreted as a sign for possibly \emph{cyclic} amplitude variations. In this context it is worth pointing out that Polaris' pulsational RV amplitude is abnormally low, even among Cepheids pulsating in the first overtone. To wit, \citet{2009A&A...504..959K} determined the average peak-to-peak amplitude of single (binary) FO Cepheids in the Milky Way to be $15.8 \pm 4.1$\,\kms ($18.2 \pm 4.7$\,\kms), with V1726\,Cyg having the lowest amplitude of $8.5$\,\kms\ (not counting Polaris).

Polaris' variable photometric amplitudes have also been studied in detail following the initial discovery of amplitude variations by \citet{1983ApJ...274..755A}. For example, \citet{2008MNRAS.388.1239S} and \citet{2008ApJ...683..433B} independently analyzed photometric observations by the Solar Mass Ejection Imager ({\it SMEI}) instrument on board the {\it Coriolis} space craft and concluded that the photometric amplitude of Polaris showed a significant increase between 2004 and 2007, thus creating an important link between photometric amplitude growth and the contemporaneous reports of growing RV amplitudes. 

Additional periodic signals were reported by \citet[$45.3$\,d]{1989AJ.....98.2249D}, \citet[$34.3$\,d]{1998AJ....116..936K}, \citet[$17.03$ and $40.2$d; henceforth: HC00]{2000AJ....120..979H}, \citet[$119$\,d]{2008AJ....135.2240L}, and \citet[$2-6$d; henceforth: B+08]{2008ApJ...683..433B}. However, none of these studies confirmed any previously reported periodicities (not even using contemporaneous data sets), and B+08 concluded that any detections of signals on periods longer than $6$\,d had been spurious and caused by instrumental drifts or complicated spectral windows. 

Meanwhile, Cepheid light curves in general have been shown to be much more complex than previously thought \citep[e.g.][]{2008AcA....58..313P,2015MNRAS.446.4008E,2015MNRAS.454..849P,2017MNRAS.464.1553D,2017MNRAS.468.4299S,2018A&A...610A..86S} and  high-precision RV observations have revealed both cycle-to-cycle and long-term modulations \citep{2014A&A...566L..10A}. Unfortunately, the relation between photometric and velocimetric variability modulations remains unclear due to a combination of observational selection effects, particularly involving brightness limits, available reference stars, and modulation timescales. Among the best-studied cases is the long-period Cepheid $\ell$~Car (\Ppuls$=35.5$d) that exhibits cycle-to-cycle RV curve modulations as well as temporal variations of its maximum angular diameter. Intriguingly, a contemporaneous study using optical/near infrared interferometric and optical RV data showed that both types of signals were modulated, yet that there was no direct correspondence between the two modulation signals \citep{2016MNRAS.455.4231A}. Hence, photospheric motions (traced by photometry and interferometry) and gas motions (traced by spectral lines) appear to be affected by different physical processes responsible for the modulations.

As the above shows, the literature on Polaris contains many conflicting reports concerning amplitude variations, some of which are based on highly inhomogeneous data sets, as well as the presence of additional periodic signals. A reconsideration of the (in-)stability of Polaris' RV variability is thus in order, in particular in light of the recent reports of a renewed decline in RV amplitude. 

This paper is structured as follows. \S\ref{sec:data} presents a new set of highly precise RVs of Polaris (\S\ref{sec:obs}) as well as an empirical template fitting method (\S\ref{sec:templates}) used to investigate long-term RV variations. \S\ref{sec:results} presents the results of applying the template fitting method to {\it Hermes} RVs (\S\ref{sec:HermesRVT}) and publicly accessible datasets from the literature (\S\ref{sec:LiteratureRVT}). \S\ref{sec:orbit} provides an updated orbital solution for the 29-year orbit of the Polaris Aa-Ab system based on the results from the template fitting method. The discussion in \S\ref{sec:discussion} focuses on reliability of reported amplitude variations and additional periodicities. Specifically, \S\ref{sec:disc:limitations} discusses the impact of data inhomogeneity on RV amplitudes, \S\ref{sec:disc:SMEI} investigates amplitude variations in the full {\it SMEI} data set, \S\ref{sec:BIS} considers additional periodic signals revealed by line bisector measurements, and \S\ref{sec:disc:rotation} aims to consolidate the information from previous and new findings to achieve a clearer picture of Polaris. The final section \S\ref{sec:conclusions} summarizes the results and conclusions.

\section{Observational Data and Analysis\label{sec:data}}

\subsection{{\it Hermes} Observations\label{sec:obs}}

\begin{figure*}
\centering
\includegraphics{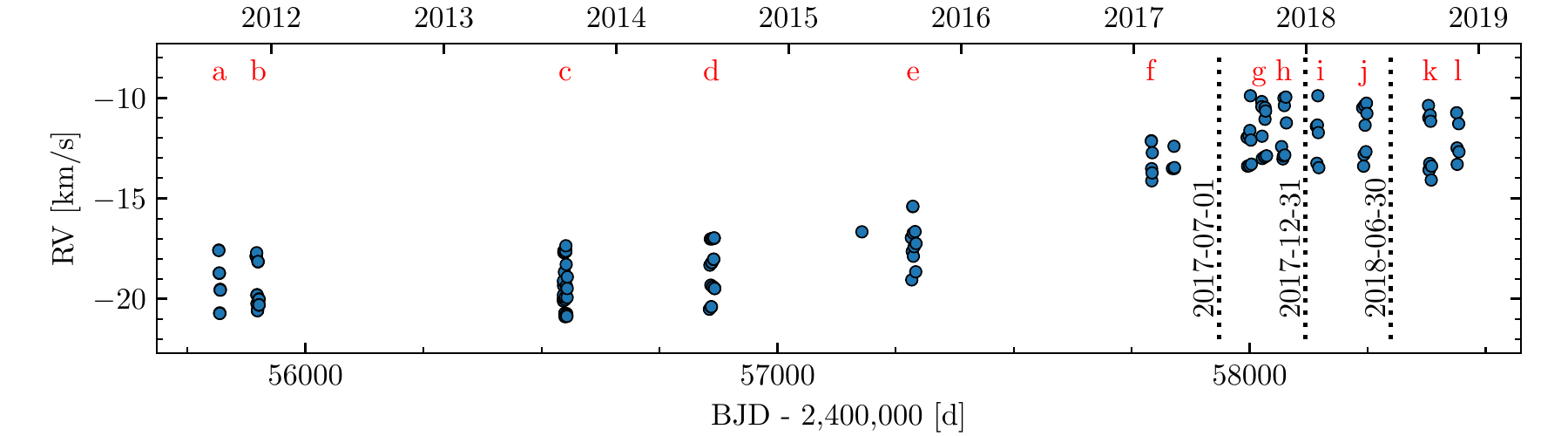}
\caption{{\it Hermes} radial velocity curve of Polaris as a function of observing date. The short term scatter is due to the $\sim 4$\,d pulsation, the long-term variations reveal the orbital motion whose pericenter passage occurred in late 2016. \label{fig:RVcurve}}
\end{figure*}

Polaris was observed using the high-resolution Echelle spectrograph {\it Hermes} as part of a large observing program dedicated to high-precision velocimetry of classical Cepheids (Anderson et al., in prep.). {\it Hermes} features a resolving power of $R \sim 85,000$ and is mounted to the Flemish 1.2m Mercator telescope located on the Roque de los Muchachos Observatory on La Palma, Canary Island, Spain \citep{2011A&A...526A..69R}. All observations were made using the high-resolution fiber to achieve the best throughput and spectral resolution. The data were reduced using the dedicated reduction pipeline that carries out standard processing steps such as flat-fielding, bias corrections, order extraction, and cosmic clipping. The measurements are provided in Appendix \ref{app:HermesRVs} as well as via the CDS\footnote{\url{https://cds.u-strasbg.fr/}}.

The 161 observations presented here were gathered between 11 September 2011 and 20 November 2018, mostly during observing runs of 9-11 nights duration. Occasionally, observations could be secured in between the regular observing runs through time exchanges with other groups. RVs are determined using the cross-correlation method \citep{1996A&AS..119..373B,2002A&A...388..632P}.
The measured RV is defined as the center of a Gaussian profile fitted to a cross-correlation function (CCF) computed using the spectral orders $55-74$ and a numerical mask that includes the position and relative strength of approximately $1130$ metallic absorption lines found in a G2 star of Solar metallicity. The median signal-to-noise (S/N) of the spectra in this wavelength  range is $230$. Further details can be found in \citet{2015ApJ...804..144A} and will be provided as part of the full Cepheid RV catalog (Anderson et al. in prep.), including derived velocities of RV standard stars \citep{1999ASPC..185..383U}. Figure \ref{fig:RVcurve} shows the full {\it Hermes} RV time series presented here.

Several steps were taken to achieve maximum short-term RV precision and track long-term RV stability. To ensure short-term precision, we have continuously monitored the nightly evolution of the ambient pressure, and re-calibrated the wavelength solution whenever pressure variations exceeded $\sim 0.4 - 0.5$\,mbar, thereby reducing intra-night variations of the instrument's RV zeropoint. Additionally, we compute RV drift corrections due to pressure changes following \citet[Ch. 2.3]{2013PhDT.......363A}. This combined procedure achieves a short-term instrumental stability of approximately $10 - 15$\ms\ over the duration of one observing run (10 nights). 

The long-term stability was monitored by means of RV standard stars. A preliminary analysis of the standard star monitoring indicates a very stable RV zeropoint (better than 20\ms) from 2011 to 2017 with a possible increase of about $50-70$\ms\ after 2017.
The investigation of this offset is ongoing.
Long-term RV zero-point variations to first order lead to changes in the mean RV at a given epoch; a spectral type-dependence of any zero-point offsets can lead to phase-dependent changes in the derived RV. However, such changes lead to (in this case) negligible second order effects on the order of a few \ms .

\subsection{RV Template Fitting \label{sec:templates}}
We determine temporal variations of the pulsation-averaged velocity, $v_\gamma$, using a self-consistent empirical template fitting approach. The method was presented in detail by \citet{2016ApJS..226...18A}, where it was used to investigate orbital motion in long-period Cepheids. In principle, the template fitting method can simultaneously solve for \Ppuls\ variations. However, the short time span of the individual observing epochs of typically $5-10$\,d ($1-2.5$ pulsation cycles), with one exception of $35$d, is too short to precisely constrain \Ppuls\ for each epoch. After experimenting with time-variable \Ppuls\ at first, we found employing a fixed value for \Ppuls\ to be a sounder approach. To this end, we first subtract an orbital model\footnote{Two different orbital models were used: the orbital solution obtained from a first pass with time-variable \Ppuls, and the orbital solution derived from a combined Keplerian and Fourier Series fit to the {\it Hermes} data initialized at the K96 orbital period. Both approaches yield \Ppuls\ consistent to within $0.25\sigma$.} from the RV time series and then determine the peak of highest power in a Lomb-Scargle periodogram \citep{1976Ap&SS..39..447L,1982ApJ...263..835S} computed on the residuals using \texttt{astropy}\footnote{\url{http://www.astropy.org}}. The uncertainty on \Ppuls\ can be determined from a $\chi^2$ distribution computed for a dense grid of input periods, so that $\chi^2_{1\sigma} = \chi^2_{\rm{min}}+1$. However, in the case of Polaris, $\chi^2_{\rm{min}} / N_{\rm{DOF}} \sim 15.4 \gg 1$ due to relatively large residual scatter (RMS$\sim 223$\ms) after the orbit removal. We therefore adopt $\chi^2_{\rm{min}}+15.4$ as the $1\sigma$ confidence level and find \Ppuls$=3.97198 \pm 0.00004$\,d at Epoch $2\,456\,553.62553$ in $2\sigma$ agreement with $3.97209 \pm 0.00004$\,d derived on {\it SMEI} data between $2003-2007$ \citep{2008MNRAS.388.1239S}. 

We adopt a reference epoch to define the pulsation template. This reference epoch is chosen from the full time series data such that good phase coverage is achieved over a relatively short timescale in order to avoid noise from period or amplitude fluctuations as well as orbital motion. The template to be applied to all other epochs is then defined by the best-fit Fourier series model (we find that two harmonics are best-suited for {\it Hermes} data) of this reference data set using \Ppuls$=3.97198$\,d determined above. The full RV time series is then sub-divided into epochs of short duration (cf. alphabetic labels in Figure \ref{fig:RVcurve}) and the reference pulsation model is fitted to each epoch using a least-squares routine that varies offsets in $v_\gamma$ and determines a phase offset such that the pulsation phase $\phi \equiv 0$ at minimum RV.

\begin{figure}[t]
\centering
\includegraphics{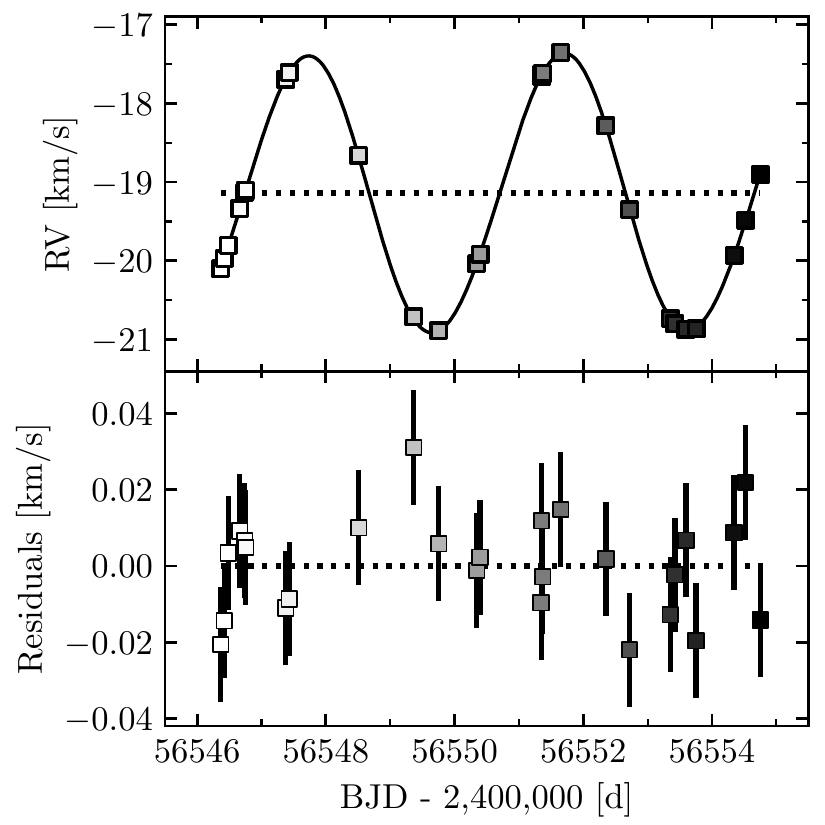}
\caption{RVs of Polaris at the reference epoch (c) chosen for its densest phase coverage over a short timescale. The top panel shows the {\it Hermes} data as a function of Barycentric Julian Date with the overplotted model consisting of the reference Fourier series model and a slow linear term of $9 \pm 1$\ms$\,\rm{d}^{-1}$. The bottom panel shows the residuals of the fit, with an RMS of $13$\ms. The grayscale traces observing date as is done for the template fits below.}\label{fig:reference_curve}
\end{figure}

The most suitable reference epoch was epoch (c) in mid 2013, which is near the minimum RV of the 29-year orbit. The reference epoch consists of 26 measurements secured over the course of 9 nights and is shown in Fig. \ref{fig:reference_curve}. The peak-to-peak amplitude is $3.526 \pm 0.006$\kms\ and Table \ref{tab:FourierRef} lists the Fourier coefficients derived. We find that a linear trend $\dot{v} = 8.7 \pm 1.1$\ms\,d$^{-1}$ is needed to account for orbital motion and possible additional RV variations (cf. \S\ref{sec:periods}) during the reference epoch. We therefore introduce an additional fit parameter, $\dot{v}$, in the template fitting routine in order to account for trends acting on the timescale of an observing epoch. 

The reference epoch's reduced $\chi^2$ is unity for an assumed short-term precision of approximately $13$\ms. We therefore adopt $15$\ms\ as the error of the {\it Hermes} RVs. This value agrees well with the short-term RV precision determined using RV standard stars, indicating that the zero-point stability is the dominant uncertainty on these measurements as previously found for $\delta$~Cep \citep{2015ApJ...804..144A}. While $15$\ms\ is a reasonable estimate of the short-term RV precision, longer-term zero-point variations affect $v_\gamma$ at the level of up to $50-70$\ms\ over the 7 year time span. Therefore, the larger error of $70$\ms\ is adopted for determining the orbit (\S\ref{sec:orbit}).

\begin{table}
\centering
\begin{tabular}{lrr}
\hline
Source & $a$ & $b$ \\ 
 & [\kms] & [\kms] \\ 
 \hline
{\it Hermes} h1 &  $-1.763 \pm 0.005$ &  $-0.005 \pm 0.005$ \\
{\it Hermes} h2 &   $0.026 \pm 0.005$ &   $0.044 \pm 0.006$ \\
K96 `08' & $-0.759 \pm 0.109$ &  $-0.002 \pm 0.106$  \\ 
K96 `CE' & $-0.813 \pm 0.012$ &  $-0.001 \pm 0.012$  \\
Gorynya & $-0.61 \pm 0.24$ &   $0.001 \pm 0.24$ \\ 
 \hline
\end{tabular}
\caption{Fit coefficients determined for the reference epoch per data set. The reference RV model is obtained as $v_r = v_\gamma + \sum_j a_j \cdot \sin{2\pi j\phi_{\rm{puls}}} + b_j \cdot \cos{2\pi j\phi_{\rm{puls}}}$. In the case of {\it Hermes}, we use two harmonics $j$ (h1,h2) and allow for an additional linear trend, $\dot{v} = \rm{d}v/\rm{d}t$, per epoch to account for orbital motion and additional periodicities on timescales longer than 10\,d. Phase $\phi$ is computed using \Ppuls$=3.97198 \pm 0.00004$\,d and $t_0 = 56\,553.62553$. For literature data, we use a single harmonic (sinusoidal) RV curve shape\label{tab:FourierRef}. Phase $\phi \equiv 0$ is set to the minimum RV, so that $a_1 < 0$ and $b_1 = 0$ for sinusoidal RV curves. We adopt \Ppuls$=3.9721$d from K96 to fit the `08', `CE', and Gorynya data sets.}
\end{table}

\section{Results \label{sec:results}}

\subsection{RV Template Fitting Applied to {\it Hermes} Data \label{sec:HermesRVT}}

\begin{figure*}
\centering
\includegraphics{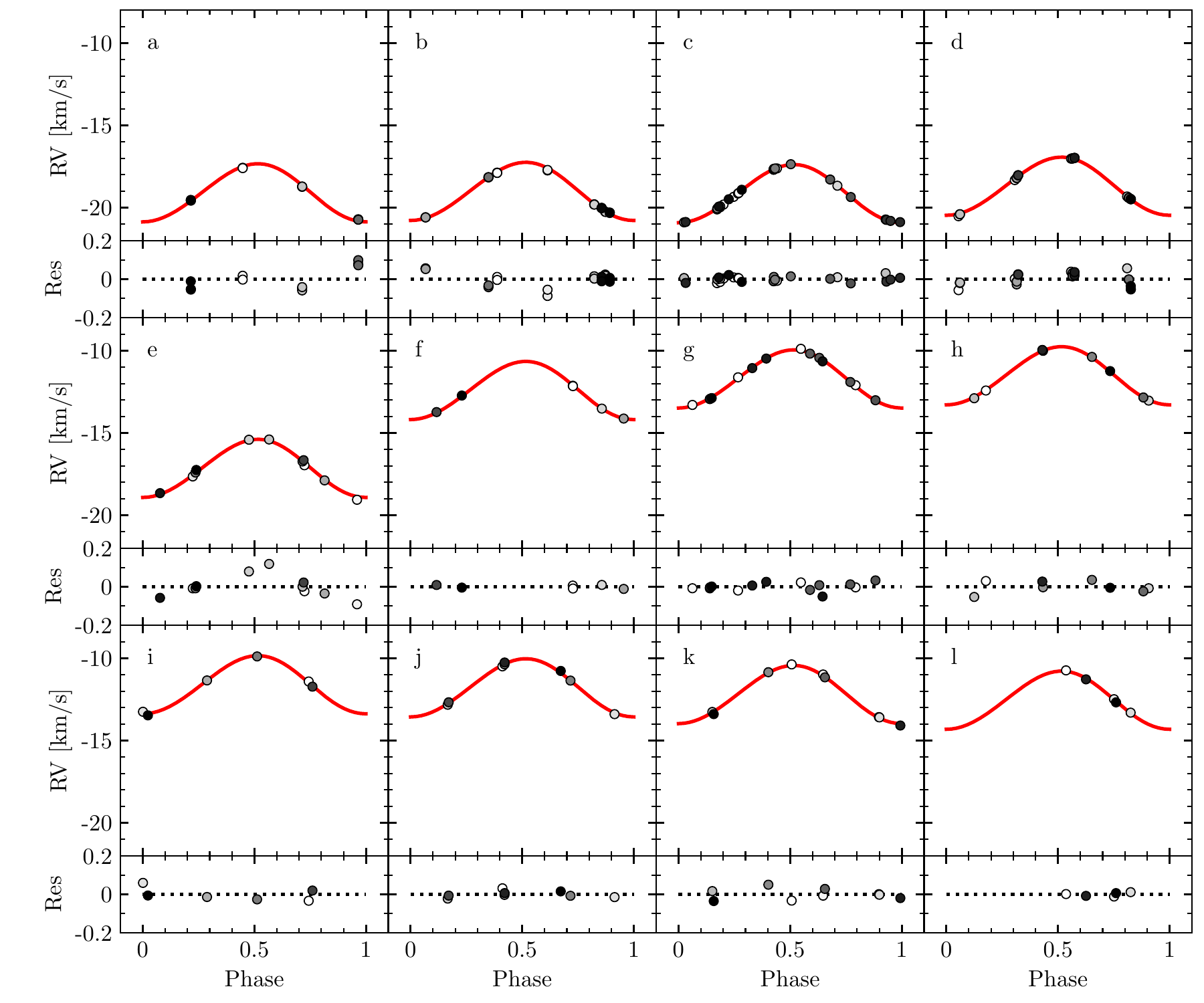}
\caption{Per-epoch template fits to {\it Hermes} RV data. In each panel, a gray scale is applied to distinguish the observing data internal to the given epoch, i.e., the gray scale runs from white to black among the measurements of that epoch. For each epoch, we show the {\it Hermes} data phase-folded and zero-phase shifted as determined from the template fit. The red solid curve in each panel is the pulsation model (cf. Fig.\,\ref{fig:reference_curve}), shifted to the $v_\gamma$ of that epoch\label{fig:templates}. Fit residuals are shown underneath each template fit, and a grayscale traces observing date in each epoch (white to black).}
\end{figure*}

The RV template fitting procedure (\S\ref{sec:templates}) solves for three parameters: $v_\gamma$, $\Delta \phi$, and $\dot{v}$ using the fixed set of Fourier coefficients determined from the reference epoch. Since the amplitude of the fitted model is constant, any temporal variations of Polaris' RV amplitude would lead to increased fit residuals, in particular near the minimum and maximum of the pulsation-induced RV variability. Assuming perfect phase coverage, a linear dependence of RV curve amplitude on time would tend to introduce a dependence of residual scatter on time. Conversely, a constant scatter in the fit residuals indicates stable RV variability.

\begin{figure*}
\centering
\includegraphics{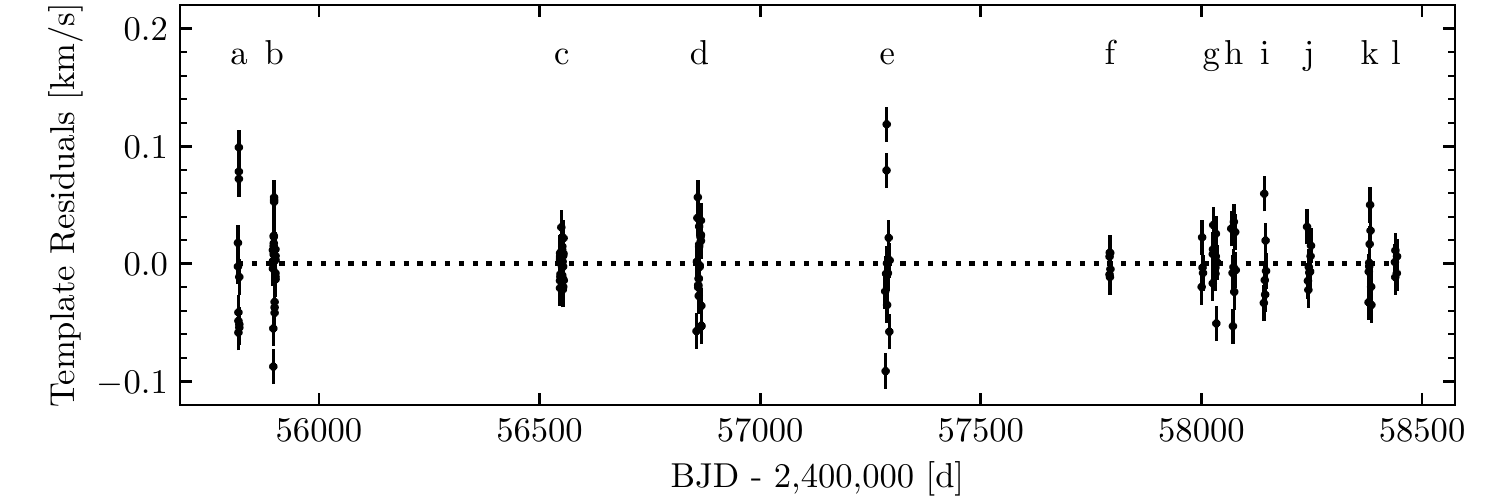}
\caption{{\it Hermes} RV template fit residuals as a function of observing date. The highest scatter is found in epoch (e), which is near pericenter passage and thus the most sensitive to orbital motion.} \label{fig:template_residuals}
\end{figure*}

The results from the template fitting applied to  {\it Hermes} RVs are illustrated in Fig.\,\ref{fig:templates} and listed in Tab.\,\ref{tab:templateresults}, including the values of $v_\gamma$, $\dot{v}$ and each epoch's fit RMS. Each panel in Fig.\,\ref{fig:templates} shows the RV data and template fit for one epoch as labeled in Fig.\,\ref{fig:RVcurve}. The data in each panel are color-coded to trace the observing date during that epoch. Figure\,\ref{fig:templates} readily shows the long-term orbital motion that slowly modulates the per-epoch average of the RV curve. Residuals are shown underneath each panel. The mean per-epoch RMS is $28$\ms, with a maximum of $56$\ms\ during epochs a and e. Epoch e is close to pericenter passage and the most sensitive to orbital motion within the duration of the epoch's observations. Figure \ref{fig:template_residuals} illustrates the template fit residuals whose minimum and maximum values are $-91$\ms\ and $+119$\ms\ (both in epoch e), with an overall mean residual of  $31$\ms. By construction, a very close fit is obtained for the reference epoch (c). Less densely sampled epochs can exhibit larger scatter due to variations or fluctuations of \Ppuls\ as well as potentially real variations in the RV curve shape. However, Figures \ref{fig:templates} and \ref{fig:template_residuals} clearly demonstrate that the pulsational RV amplitude of Polaris was stable at the level of $\sim 30$\ms\ over the entire duration of the observing program (2011-2018). Specifically, they exclude the large, \kms-level amplitude variations between 2017-2018 (traced by epochs g through j) reported by \citet{2018MNRAS.481L.115U}. 

\begin{table*}
\centering
\begin{tabular}{lrrrrrrrrr}
\hline
Epoch & $N_{\rm{obs}}$ & Mean BJD & Min BJD & Max BJD & $v_\gamma$ & $\sigma_{v_{\gamma}}$ & $\dot{v}$ & $\sigma_{\dot{v}}$ & RMS \\
 & & & & &  [\kms] & [\kms] & [\kms$\rm{d^{-1}}$] & [\kms$\rm{d^{-1}}$] & [\kms] \\
\hline
\multicolumn{8}{c}{{\it Hermes} RVs, \Ppuls$=3.97198$d, peak-to-peak amplitude $3.526$\kms}\\
\hline
a & 11 & 55818.23 & 55816.54 & 55819.60 & -19.119 & 0.021 &  0.0576 & 0.0219 & 0.056\\
b & 23 & 55898.72 & 55895.77 & 55901.75 & -19.034 & 0.008 &  0.0283 & 0.0039 & 0.034\\
c$^\dagger$ & 26 & 56550.43 & 56546.37 & 56554.75 & -19.168 & 0.003 &  0.0087 & 0.0010 & 0.013\\
d & 25 & 56862.22 & 56855.73 & 56866.74 & -18.719 & 0.007 &  0.0145 & 0.0022 & 0.030\\
e & 11 & 57288.02 & 57283.41 & 57293.40 & -17.177 & 0.021 &  0.0278 & 0.0071 & 0.056\\
f & 6  & 57792.63 & 57791.80 & 57793.80 & -12.448 & 0.015 &  0.0442 & 0.0531 & 0.009\\
g & 14 & 58022.23 & 58000.49 & 58035.77 & -11.748 & 0.006 & -0.0039 & 0.0005 & 0.021\\
h & 8  & 58073.06 & 58067.64 & 58077.79 & -11.555 & 0.014 &  0.0109 & 0.0053 & 0.028\\
i & 6  & 58143.95 & 58141.39 & 58146.47 & -11.639 & 0.023 & -0.0422 & 0.0131 & 0.032\\
j & 8  & 58244.15 & 58239.38 & 58248.37 & -11.826 & 0.008 &  0.0243 & 0.0029 & 0.016\\
k & 9  & 58381.75 & 58378.77 & 58385.32 & -12.234 & 0.012 & -0.0333 & 0.0059 & 0.026\\
l & 5  & 58440.69 & 58438.45 & 58443.32 & -12.578 & 0.023 & -0.0262 & 0.0048 & 0.009\\
\hline
\multicolumn{10}{c}{High-dispersion (8 \AA/mm) photographic RVs (K96-08), \Ppuls$=3.9721$d, $2K = 1.518$\kms}\\
\hline
a           & 34 & 45526.29 & 45414.53 & 45592.51 & -19.555 & 0.218 & $-$ & $-$ & 1.189\\
b           &  6 & 45890.18 & 45852.85 & 45920.67 & -19.355 & 0.701 & $-$ & $-$ & 1.213\\
c           & 39 & 46247.84 & 46184.55 & 46290.56 & -17.972 & 0.085 & $-$ & $-$ & 0.486\\
d           & 12 & 46624.87 & 46575.77 & 46724.88 & -16.046 & 0.180 & $-$ & $-$ & 0.538\\
e           & 27 & 46964.16 & 46944.61 & 46993.65 & -14.065 & 0.163 & $-$ & $-$ & 0.789\\
f           & 10 & 47306.51 & 47299.86 & 47309.82 & -12.221 & 0.162 & $-$ & $-$ & 0.419\\
g$^\dagger$ & 13 & 47595.82 & 47582.65 & 47612.81 & -11.808 & 0.153 & $-$ & $-$ & 0.459\\
h$^\dagger$ & 24 & 47653.28 & 47628.62 & 47674.75 & -12.106 & 0.098 & $-$ & $-$ & 0.442\\
i$^\dagger$ & 16 & 47700.48 & 47675.72 & 47724.70 & -12.257 & 0.178 & $-$ & $-$ & 0.601\\
\hline
\multicolumn{10}{c}{RVs from digitally recorded DDO spectra (K96-CE), \Ppuls$=3.9721$d, $2K = 1.626$\kms}\\
\hline
a           & 23 & 48843.93 & 48828.67 & 48850.69 & -13.921 & 0.078 & $-$ & $-$ & 0.328\\
b           & 19 & 49021.73 & 49017.52 & 49023.92 & -14.989 & 0.027 & $-$ & $-$ & 0.107\\
c           & 26 & 49061.43 & 49058.58 & 49064.75 & -15.097 & 0.037 & $-$ & $-$ & 0.176\\
d           & 21 & 49452.62 & 49450.65 & 49454.59 & -15.420 & 0.016 & $-$ & $-$ & 0.068\\
e           & 13 & 49479.27 & 49476.60 & 49481.56 & -15.488 & 0.016 & $-$ & $-$ & 0.044\\
f           & 15 & 49564.88 & 49563.73 & 49565.80 & -15.708 & 0.037 & $-$ & $-$ & 0.123\\
g$^\dagger$ & 27 & 49599.87 & 49597.59 & 49601.87 & -15.926 & 0.008 & $-$ & $-$ & 0.040\\
\hline
\multicolumn{10}{c}{{\it Coravel}-type Gorynya RVs, \Ppuls$=3.9721$d, $2K = 1.22$\kms}\\
\hline
a           & 25 & 49468.49 & 49446.34 & 49488.32 & -16.217 & 0.096 & $-$ & $-$ & 0.435\\
b$^\dagger$ & 15 & 49566.89 & 49564.54 & 49569.57 & -16.240 & 0.162 & $-$ & $-$ & 0.627\\
\hline
\end{tabular}
\caption{Results from the template fitting analysis applied to RV data of Polaris from {\it Hermes}, \citet{1996JRASC..90..140K}, and \citet{1992SvAL...18..316G}. $^\dagger$ labels the reference epoch(s) used for defining the RV template for each data set. No corrections for instrumental zero-point differences or variations have been applied in this table. \label{tab:templateresults}}
\end{table*}

\subsection{RV Template Fitting Applied to Literature Data \label{sec:LiteratureRVT}}

\begin{figure}
\centering
\includegraphics{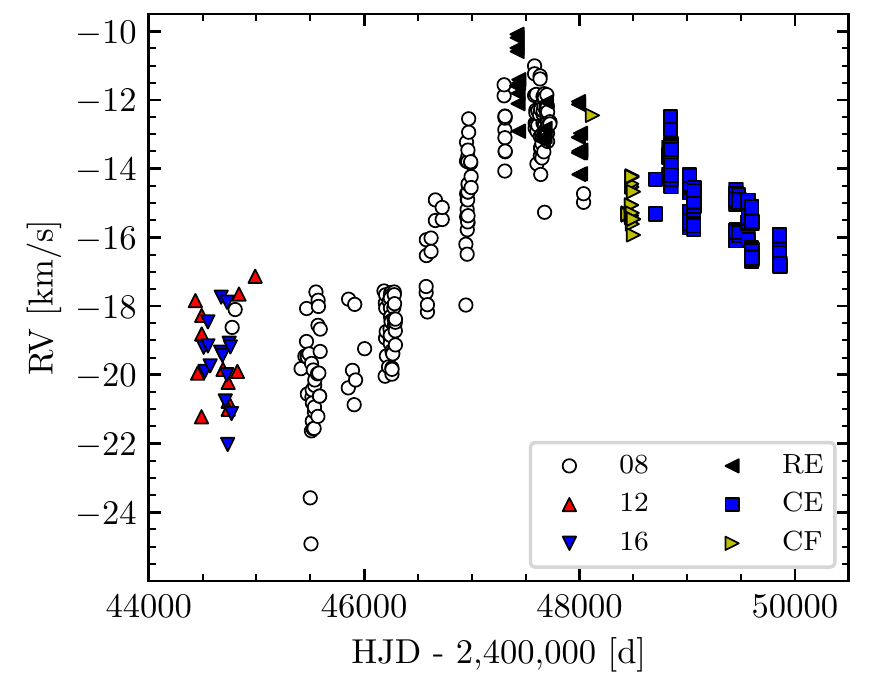}
\caption{Radial velocity data from \citet{1996JRASC..90..140K} measured at DDO. Different symbols and colors are used to distinguish data measured using different instrumental setups. Labels `08', `12', and `16' refer to photographic of different dispersions (in mm/\AA). `RE' refers to measurements taken using a Reticon detector, `CF' to ones using a fiber link, and `CE' to observations made using a CCD. Instrumental zero-point variations related to these interventions and associated calibration issues are readily apparent, especially following the peak of the orbital RV variation near HJD 2\,445\,750\label{fig:Kamper1996_all}}
\end{figure}

We apply the RV template fitting method to literature RVs in order to test RV curve stability for separate, homogeneous data sets and to determine the temporal variation of $v_\gamma$. To this end, we employ: RVs measured by \citet[][henceforth: K96]{1996JRASC..90..140K} using high-dispersion (8\,mm/\AA) photographic DDO spectra (K96-08 RVs), RVs measured by K96 using the DDO spectrograph following the installation of a 1024x1024 Thomson CCD (K96-CE RVs), and RVs determined by \citet{1992SvAL...18..316G} using the Moscow {\it Coravel}-type correlation spectrometer \citep{1987SvA....31...98T} (Gorynya RVs). The  K96-08 and K96-CE datasets are the two largest subsets of the data published by K96. As Figure\,\ref{fig:Kamper1996_all} shows, various interventions at the telescope led to noticeable zero-point changes, which were occasionally accompanied by significant spurious zero-point fluctuations, e.g. following the installation of the CCD. Specifically, K96 noted that `observations were standardized' only \emph{after} the ``surprising decrease in pulsational amplitude'' by \citet{1983ApJ...274..755A}. The K96-CE data are internally more precise on short timescales than the K96-08 data, although the template fits indicate calibration issues in the first season after the CCD camera had been installed at the spectrograph. Unfortunately, the zero-point seems to vary significantly between each observational epoch (cf. Fig.\ref{fig:Kamper1996_all}, blue squares), so that  the inferred values of $v_\gamma$ are not useful for determining the orbit.

The Gorynya RVs are relatively few and noisy compared to the K96 data sets. However, the larger catalog of Cepheid RVs published by \citet{1992SvAL...18..316G,1996AstL...22..175G,1998AstL...24..815G} provides a unique and fairly homogeneous basis for comparing Cepheid RVs over timescales of several decades. Based on a preliminary analysis of Cepheids observed as part of our ongoing program (Anderson et al in prep.), the Gorynya data generally agree very well with RVs determined using {\it Hermes}. For instance, we determine peak-to-peak RV amplitudes of $27.70$\,\kms\ vs. $27.73$\,\kms\ for $\zeta$~Gem (using 5 harmonics), and $17.11$\,\kms\ vs. $17.24$\,\kms\ for EU~Tau based on {\it Hermes} and Gorynya data, respectively. Other RV datasets previously discussed in the literature were either not publicly available, or were not listed on an absolute scale so that the orbital motion cannot be inferred directly.

Table\,\ref{tab:FourierRef} lists the Fourier coefficients of the sinusoidal reference curves and Tab.\,\ref{tab:templateresults} lists the results of the template fits applied to literature RVs. Given the  lower precision of the literature data, we do not solve for any linear trends $\dot{v}$ as was the case for the {\it Hermes} data. Appendix \ref{app:Kamper}  illustrates the template fit results of the K96-08 and K96-CE data. None of the three literature data sets analyzed here exhibit noticeable RV amplitude differences exceeding a few hundred \ms, i.e., at the level of the precision of the data. Instead, the reference epochs of the three literature data sets yield stable peak-to-peak RV amplitudes that agree to within their uncertainties: $1.51 \pm 0.21$\kms\ (K96-08), $1.63 \pm 0.02$\kms\ (K96-CE), and $1.2\pm 0.4$\kms\ (Gorynya). The aforementioned general agreement between a large number of Cepheid RV amplitudes based on Gorynya's data and {\it Hermes} would tend to support the interpretation that the pulsational RV variability of Polaris has changed over time. Specifically, this would imply  that the RV amplitude more than doubled between 1995 and 2011, while homogeneous data sets recorded both before and after this period exhibit no significant changes. However, we notice that many of the Gorynya RVs exhibit scatter of up to $1-2$\kms\ among measurements obtained in rapid succession (on timescales of minutes), which could point to a problem with these particular measurements.

For comparison, \citet{2008AJ....135.2240L} reported non-monotonous variations in RV amplitude with  $2K = 2.2$, $2.1$, and $2.4$\,\kms\ in 2005, 2006, and 2007, whereas B+08 (cf. their Fig.\,4) reported a rather monotonous increase in amplitude from $2K \sim 1.9$ to $2.1$ and $2.3$\,\kms\ during the same years. B+08 derived a linear relation for the RV amplitude  of $A_{\rm{RV}}(t) = (0.90 \pm 0.01) + (1.45 \pm 0.15) \times 10^{-4}(t-t_0)$\kms, where $t_0 = 2\,453\,000$. Converting this relation to peak-to-peak amplitudes ($2K$) and projecting it to the first epoch of {\it Hermes} measurements would imply $2K \sim 2.62$\kms, which is significantly less than the $2K = 3.5$\kms\ observed using {\it Hermes}, cf. Tab.\,\ref{tab:templateresults}. Taken at face value, this implies that the RV amplitude has increased much faster between 2007 and 2011 than between 2004 and 2007, and that it has since remained nearly constant.  Calculating the amplitude expected for the last K96-CE epoch using this linear relation yields $2K \sim 0.81$\kms, which is much lower than the observed $1.6$\kms. Hence, the comparison of results based on data sets with differing systematics suggests that Polaris' RV amplitude may vary significantly in a non-linear fashion over timescales of up to a few years. 
Given the strong interest in Polaris, it is somewhat surprising and very unfortunate that there exists no homogeneous dataset that unambiguously demonstrates the fast and non-linear RV amplitude variations implied by this comparison of literature amplitudes, especially since it is difficult to distinguish between (small) amplitude changes caused by astrophysical effects and non-astrophysical systematics related to drifting RV zero-points, inhomogeneous data sets, or incompatible RV measurement definitions, cf. \S\ref{sec:disc:limitations}.

\subsection{An Updated Orbital Solution for Polaris Aa-Ab \label{sec:orbit}}

\begin{table}
\centering
\begin{tabular}{@{}llcc@{}}
\hline
Element & Units & \citet{1996JRASC..90..140K} & This work \\
\hline 
$P_{\rm{orbit}}$    & [yr]        &  $29.59 \pm 0.02$   &  $29.32   \pm  0.11 $ \\
$e$                 &             &  $0.608 \pm 0.005$  &  $0.620   \pm  0.008$\\
$T_0$               & [yr]        &  $1928.48 \pm 0.08$ &  $2016.91 \pm  0.10$\\
$v_\gamma$          & [\kms]      &  $-16.42 \pm 0.03$  &  $-15.387 \pm  0.040$\\
$K$                 & [\kms]      &  $3.72 \pm 0.03$    &  $3.768   \pm  0.073$\\
$\omega$            & [deg]       &  $303.01 \pm 0.75$  &  $307.2   \pm  2.5$\\
$a\sin{i}$          & [au]        &  $2.90 \pm 0.03$    &  $2.910   \pm  0.062$\\
$f_{\rm{mass}}$     & [M$_\odot$] &                     &  $0.0286  \pm  0.260$\\
\hline
\end{tabular}
\caption{Orbital solution based on the time series of the pulsation averaged velocities listed in Tab.\ref{tab:templateresults}. The solution defined by the new {\it Hermes} and K96-08 RVs compares well to the literature solution provided by \citet{1996JRASC..90..140K}.\label{tab:OrbitResults}}
\end{table}

\begin{figure}
\centering
\includegraphics{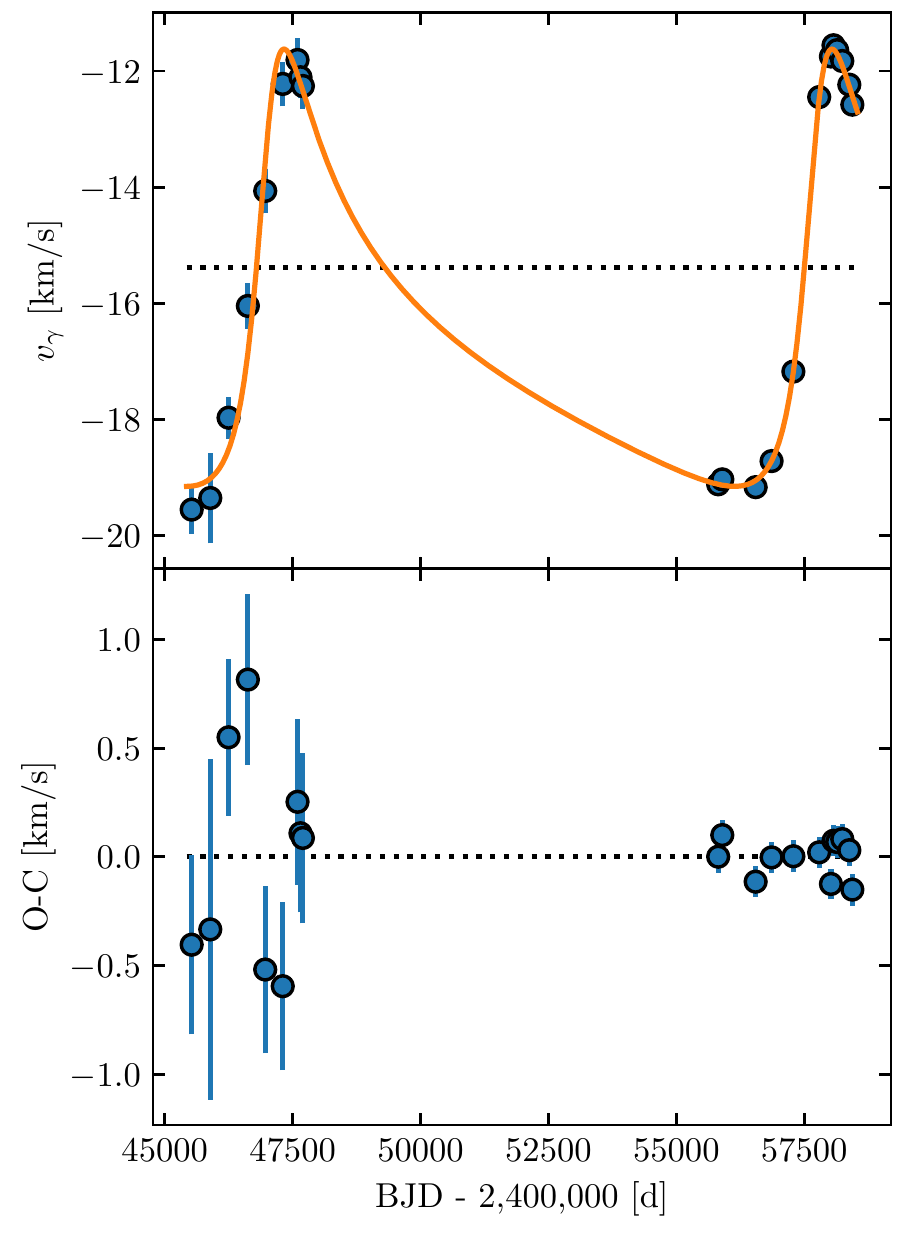}
\caption{Orbital solution based on the RV template fitting procedure applied to data from {\it Hermes} and the high-dispersion photographic RVs from \citet[K96-08 RVs]{1996JRASC..90..140K}. The K96-08 RVs have been zero-point corrected by $0.63$\kms, cf. \S\ref{sec:orbit}.\label{fig:orbitalsolution}}
\end{figure}

We determine an updated orbital solution for the Polaris Aa-Ab system using the results from the RV template fits\footnote{NB: Combining $v_\gamma$ values determined from literature data and {\it Hermes} is much more straightforward than comparing RV amplitudes related to pulsations, since orbital RV amplitudes do not depend on the spectral lines used in the measurement.} performed in \S\ref{sec:HermesRVT} and \ref{sec:LiteratureRVT}. To measure the orbit accurately, it is important to account for RV zero-point differences among the inhomogeneous literature. Combining all recent data sources is unfortunately not advisable due to a combination of insufficient knowledge of standard star velocities, unstable zero-points, and literature RVs having occasionally been corrected for orbital motion. Fortunately, both the {\it Hermes} data and the K96-08 data span the pericenter passage of the eccentric orbit as well as the extremes of the orbital RV curve. 
Correcting for the mid-point difference in the orbital RV curves, we find an approximate zero-point offset of $-0.7$\kms\ between K96-08 and {\it Hermes}. Adding $0.7$\kms\ to the K96-08 $v_\gamma$ values, we determine best-fit orbital solutions for a range of small zero-point shifts near this value. The solution with the minimum $\chi^2$ is adopted as the correct zero-point shift and yields $-0.63$\kms.
The Gorynya RVs are not used in the orbital fit, since they span a short a time interval (98\,d) and exhibit large short-term (timescale of minutes) scatter on the order of 1\kms. 

We fit a Keplerian model to the zero-point corrected $v_\gamma$ time series using \Porb$=29.6$\,yr (K96) as the starting value and determine the solution listed in Tab.\,\ref{tab:OrbitResults}. Figure\,\ref{fig:orbitalsolution} illustrates the result obtained, which for the most part agrees to within the uncertainties with the solution by K96. Besides a small difference in \Porb, the principal difference between the two solutions is the center-of-mass velocity of the binary system, which directly depends on the instrumental zero-point. The slightly larger uncertainties of our result are primarily due to the higher precision of the RV data, which are very sensitive to imperfections in the removal of the pulsation model. These imperfections lead to an elevated $\chi^2_{\rm{red}} = 2.0$, which linearly affects the reported fit uncertainties derived from the diagonal elements of the fit covariance matrix. 

Despite the slightly larger uncertainties, our new orbital solution benefits from a much higher degree of data homogeneity than previous solutions as well as the overall high accuracy of the {\it Hermes} RV zero-point. This is particularly important when considering the center-of-mass velocity of the Polaris Aa-Ab system ($-15.387 \pm 0.040$\kms), which agrees to within $1\sigma$ with Polaris B's RV of $-22.25 \pm 8.11$\kms\ as reported in {\it Gaia} DR2 \citep{2018A&A...616A...1G,2018A&A...616A...5C,2018arXiv180409372K}.

\section{Discussion \label{sec:discussion}}

\subsection{Data Inhomogeneity and Amplitude Variations \label{sec:disc:limitations}}

The first studies of Polaris' RV signals employed photographic spectra taken with the same spectrograph. \citet{1965ApJ...141.1415R} explained that observations between 1896 and January 1903 were centered on H$_\gamma$ (``old'' Mills setup), whereas observations between  summer 1903 and the last observation in 1958 were centered near $4500$\AA\ (``new'' Mills setup). Telescope flexure corrections of up to $\pm 0.3$\kms\ were included after 1920. Unfortunately, the original publications do not specify which spectral lines were measured, nor how RV was defined, e.g. whether line barycenters, line cores, or other ways of measuring line positions were used. It is also not possible to tell whether measurements were performed on multiple lines and how such measurements were averaged. For instance, K96 notes that DDO velocities were based on a cross-correlation, although it is not specified which lines were combined in this cross-correlation. This creates an issue for comparing literature amplitudes among one another, because the hydrodynamics of a Cepheid's atmosphere introduce phase shifts and amplitude differences between different transitions formed at different heights \citep[e.g.][]{1969PASP...81..732G,1993ApJ...415..323B,2005MNRAS.362.1167P,2007A&A...471..661N,2016MNRAS.463.1707A}. 

The greatly extended wavelength ranges offered by modern Echelle spectrographs connected to CCD detectors have  potentially introduced a systematic offset for RV amplitudes of pulsating stars compared to amplitudes measured based on more restricted wavelength intervals. Amplitude differences among inhomogeneous RV data sets may therefore to some extent be explained by different weighting of differentially moving atmospheric layers. In extreme cases, amplitude differences between spectral lines or measurement definitions can exceed a few \kms, cf. \citet{2018pas6.conf..193A} for a recent review.
Since the 1980s, researchers were faced with rapidly evolving instrumental technology while RV measurements became more and more common with time. This has lead to a highly inhomogeneous data set of relatively short time intervals as can be seen from the data presented by K96 as shown Fig.\,\ref{fig:Kamper1996_all}.
To what extent such layer-averaging effects can introduce amplitude variations for Polaris' generally low amplitude \citep[cf.][]{1990ApJ...351..606K,2016ApJS..226...18A} will be investigated in detail in future work. 

The following RV amplitudes ($K$ denotes semi-amplitude of a sinusoid) were previously reported using a variety of spectrographs covering different wavelength ranges: $2K=1.50 \pm 0.08$\kms\ in 1987-1988 \citep[$\lambda 4220-4700$\AA]{1989AJ.....98.2249D}; $2K \sim 1.53$\kms\ in 1994 \citep[$\Delta \lambda \sim 23.6$\AA\ centered on $\lambda 5520$\AA\ and using an iodine absorption cell]{2000AJ....120..979H}; $2K \sim 1.6$\kms\ in 1994-1997 \citep[wavelength center changed to $\lambda 6290$\AA\ to enable telluric line corrections]{1998AJ....116..936K}; $2K \sim 2.1 - 2.4$\kms\ between 2004 and 2007 \citep[][typically 160-180 lines over an unclear spectral range, between $\lambda 3600-10500$\AA, using an iodine absorption cell]{2008AJ....135.2240L}; $2K \sim 2$\kms\ with a possible time-dependence from late 2003 to late 2007 \citep[][based on 74 \emph{strong} mostly FeI lines between $\lambda 5000 - 7100$\AA]{2008ApJ...683..433B}. 

Given the issue of inhomogeneity, the cleanest evidence for RV amplitude variations is the contemporaneous detection of RV amplitude variations (cf. \S\ref{sec:LiteratureRVT}) using independent datasets by \citet{2008AJ....135.2240L} and B+08. Although B+08 provide a linear relation for the RV amplitude increase, their Fig.\,4 may indicate non-linearity in these RV amplitude changes. However, it is suspicious that such non-linearities would only occur near both ends of the time series. Moreover, although the results from the template fitting routine applied in \S\ref{sec:LiteratureRVT} confirm individually different RV amplitudes based on different data sets, we find no evidence for amplitude variations within any given homogeneous data set. Finally, the new {\it Hermes} RVs exclude significant ($>30$\ms) amplitude variations over the course of 7 years, cf. \S\ref{sec:HermesRVT}. 

To summarize, amplitude differences of $1-2$\,\kms\ have been noted among different literature data sets spanning several decades. Unfortunately, the most significant RV amplitude changes seem to have occurred \emph{in between the various studies}, i.e., when no RV observations were carried out. The cleanest examples of varying RV amplitudes on the order of $200-300$\ms\ were provided by \citet{2008AJ....135.2240L} and B+08, although these relatively weak variations do not explain the extent of the long-term variations as a whole. Hence, even more significant amplitude variations would have had to occur on relatively short timescales of at most a few years in order to explain the observed trends. However, the fact that none of the more precise, long-term RV programs\hbox{---}including K96-CE, B+08, \citet{2008AJ....135.2240L}, and the {\it Hermes} data\hbox{---}reveal such fast and significant amplitude variations casts some doubt on the assumed astrophysical origin of the reported amplitude differences.

\subsection{Anti-correlation between Amplitude and Mean Magnitude Trends in {\it SMEI}\label{sec:disc:SMEI}}

The Solar Mass Ejection Imager ({\it SMEI}) on-board the {\it Coriolis} spacecraft monitored nearly every point in the sky every 102 minutes between January 2003 and late September 2011 \citep{2003SoPh..217..319E}. {\it SMEI} thus provided extremely dense time-series photometry of very bright stars down to approximately $R \sim 10$\,mag \citep{2006ApJ...637..880B}. After about 3-4 years of observations, {\it SMEI} data were used to report on Polaris' increasing amplitude\footnote{all amplitudes mentioned are peak-to-peak amplitudes} of $\sim 1.87 \pm 0.09$\,mmag\,yr$^{-1}$ (B+08) and $1.39 \pm 0.12$\,mmag\,yr$^{-1}$ \citep{2008MNRAS.388.1239S}. These detections have played a crucial role in establishing confidence in contemporaneous RV amplitude variations reported in the literature.

{\it SMEI} continued observing Polaris until 17 days after our {\it Hermes} observations began. To the best of our knowledge, the full {\it SMEI} Polaris data have not been presented in a refereed publication. However, \citet{2515-5172-1-1-39} briefly describe results from a wavelet analysis applied to the full {\it SMEI} time series, confirming the previously reported amplitude growth at a rate of $2.40 \pm 0.08$\,mmag\,yr$^{-1}$.  \citet{2515-5172-1-1-39} further note that no significant variations of the dominant pulsation period are found, with $\dot{P} = 35 \pm 28$\,s\,yr$^{-1}$. For comparison, \citet{2010vsgh.conf..207B} reported detections of period fluctuations in four bright Cepheids observed by {\it SMEI}.

The $8.65$\,yr baseline of {\it SMEI} observations offers a unique opportunity to bridge the gap  between the RV observations reported B+08 and the new {\it Hermes} measurements since photometric amplitudes generally scale with RV amplitudes \citep[e.g.][]{2009A&A...504..959K}. To this end, we retrieved the full time series of Polaris (spanning January 2003 to 28 September 2011) from the online {\it SMEI} archive\footnote{\url{http://smei.ucsd.edu/new_smei/index.html}}. We perform a conservative and rudimentary data cleaning to remove artifacts of obviously instrumental origin as follows. First, we compute residuals of a linear fit to the time series in order to be insensitive to an obvious temporal increase in mean magnitude seen for most stars observed with {\it SMEI}. Second, we reject data points lying farther than 0.05\,mag from the detrended residuals, which is comfortably larger than the immediately apparent pulsation amplitude. Third, we invoke a proximity criterion that selects only observations whose preceding and successive detrended residuals do not differ by more than $0.01$\,mag. This is justified by the high cadence of {\it SMEI} observations and the low photometric amplitude of Polaris. Finally, we reject a group of obvious outliers observed more than 1250\,d before the mean observing date whose magnitude deviates by at least $-0.022$\,mag (a full peak-to-peak amplitude) from the detrended residuals. In total, this cleaning procedure removed approximately $10\%$ of the initial {\it SMEI} data points as outliers.

To investigate possible variability modulations as a function of time, we split the (non-detrended) cleaned time series data into 100 subsets of equal duration ($31.60$\,d) and fit a second order Fourier series models to each subset, simultaneously solving for mean magnitude and Fourier coefficients, based on which we determine peak-to-peak amplitudes. To this end, we adopt the constant value of \Ppuls$=3.97242$\,d indicated by the highest periodogram peak. For comparison, \citet{2008MNRAS.388.1239S} determined \Ppuls=$3.97209\pm 0.00004$\,d based on {\it SMEI} data spanning 2003-2007. We repeated the analysis with differing numbers of subsets and subset durations to ensure the reliability of the results. 

\begin{figure}
\centering
\includegraphics{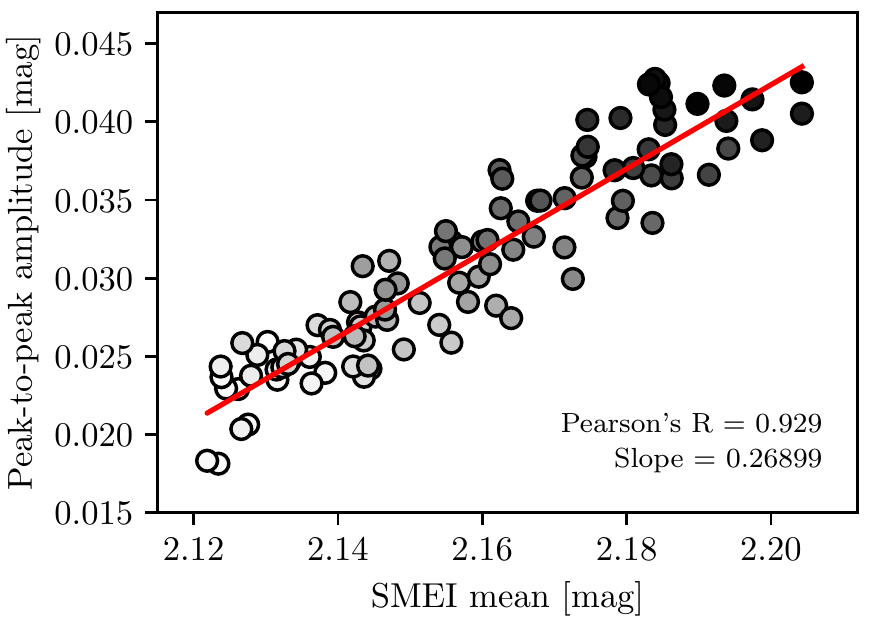}
\caption{Correlation between Polaris' growing photometric amplitude and mean magnitude as measured by {\it SMEI}. The observations cover a total baseline of 8.65 yr, lasting until 17 days after the first spectroscopic observations were made with {\it Hermes}. Observing date is color coded in grayscale to illustrate the increasing amplitude and mean magnitude with time. The red solid line is a least squares fit to the data with the slope and Pearson's correlation coefficient as indicated. \label{fig:SMEI_MeanAmp}}
\end{figure}

Figure \ref{fig:SMEI_MeanAmp} illustrates the results obtained after dividing the {\it SMEI} data into 100 subsets. There is a highly significant correlation (Pearson's $R=0.929$) between the mean magnitudes and peak-to-peak amplitudes determined. The  grayscale applied further shows that both mean magnitude and amplitude increase linearly by $0.00836$\,mag\,yr$^{-1}$ and $0.00249$\,mag\,yr$^{-1}$ \citep[consistent with][]{2515-5172-1-1-39}, respectively, over the entire $8.65$\,yr baseline of {\it SMEI} observations. Figure\,\ref{fig:SMEIepochfits} in Appendix\,\ref{app:SMEI} illustrates the Fourier series fits to each of the 100 sub-epochs. 

Our reanalysis of the {\it SMEI} data leads to several surprising realizations. First, there is no straightforward astrophysical explanation for the correlation between amplitude and mean magnitude\footnote{A companion hypothesis involving the secular evolution of an unseen companion is easily ruled out based on the slope of the correlation}. Previous studies of {\it SMEI} observations likely overlooked this correlation because of detrending algorithms applied to the data. Second, the {\it SMEI} photometric amplitude increased  linearly over the full $8.65$\,yr timespan, which includes {\it Hermes} epoch (a). It seems rather unlikely that such a long-lasting amplitude increase of astrophysical origin should have all of a sudden stopped at the beginning of the {\it Hermes} observations, which rule out RV amplitude variations at the level of $1.7\%$ ($< 30$\ms\ for $K = 1.763$\kms, cf. Tab.\,\ref{tab:FourierRef}) over more than 7 years immediately following the {\it SMEI} observations.  
Third, a  preliminary inspection of {\it SMEI} data for $\delta$\,Cep, $\eta$\,Aql, and $\zeta$\,Gem reveals correlations of similar scale and sign, albeit with larger scatter due to insufficient removal of obvious instrumental effects and larger gaps in the time series.

Pending a more detailed investigation of {\it SMEI}'s systematic uncertainties, we therefore caution that the previously reported amplitude increases based on {\it SMEI} data may be of instrumental origin. For instance, a change of the detector's non-linearity properties could lead to a simultaneous increase in amplitude and mean magnitude.

\subsection{On Additional Periodicities \label{sec:periods}\label{sec:BIS}}
Earlier studies reported the presence of additional periodicities based on RV data, although no agreement among the reported periods was established. Specifically, periods identified included signals at $17$d, $34$d, $40$d, $45$d, and $119$d \citep{1989AJ.....98.2249D,2008AJ....135.2240L,2008ApJ...683..433B}. B+08 further reported signals on timescales around $2-6$\,d based on {\it SMEI} photometry and constrained additional RV signals with periods $3-50$\,d to an upper limit $100$\ms.

The main difficulty in using RV data for a periodogram analysis is the removal of the incompletely sampled large amplitude orbital motion, especially in the case of the {\it Hermes} data that span pericenter passage. Despite this difficulty of searching for additional RV signals, we found that including slow linear trends of up to a few tens of \ms\ significantly improved the template fit results for {\it Hermes} data, cf. \S\ref{tab:templateresults}. We interpret these trends as compensating for a combination of orbital acceleration and RV signals with periods longer than the typical observing run of approximately $10$\,d. As expected by this interpretation, we find the weakest trend of $-3.9$\,\ms\,d$^{-1}$ during epoch (g), which features by far the longest time span of $35$\,d (all other epochs cover $< 11$\,d). Hence, the $40$\,d signal previously identified by \citet[henceforth: HC00]{2000AJ....120..979H} based on spectral line bisector variations in two different transitions (Sc\,II $\lambda 5527$\AA\ and Mg\,I $\lambda 5528$\AA) could potentially explain the per-epoch trends observed in {\it Hermes} RVs. We further note that the magnitude of the per-epoch trends ($\le 57.6$\,\ms\,d$^{-1}$) is roughly consistent with the upper limit of $100$\ms\ determined by B+08.

\begin{figure*}
\centering
\includegraphics{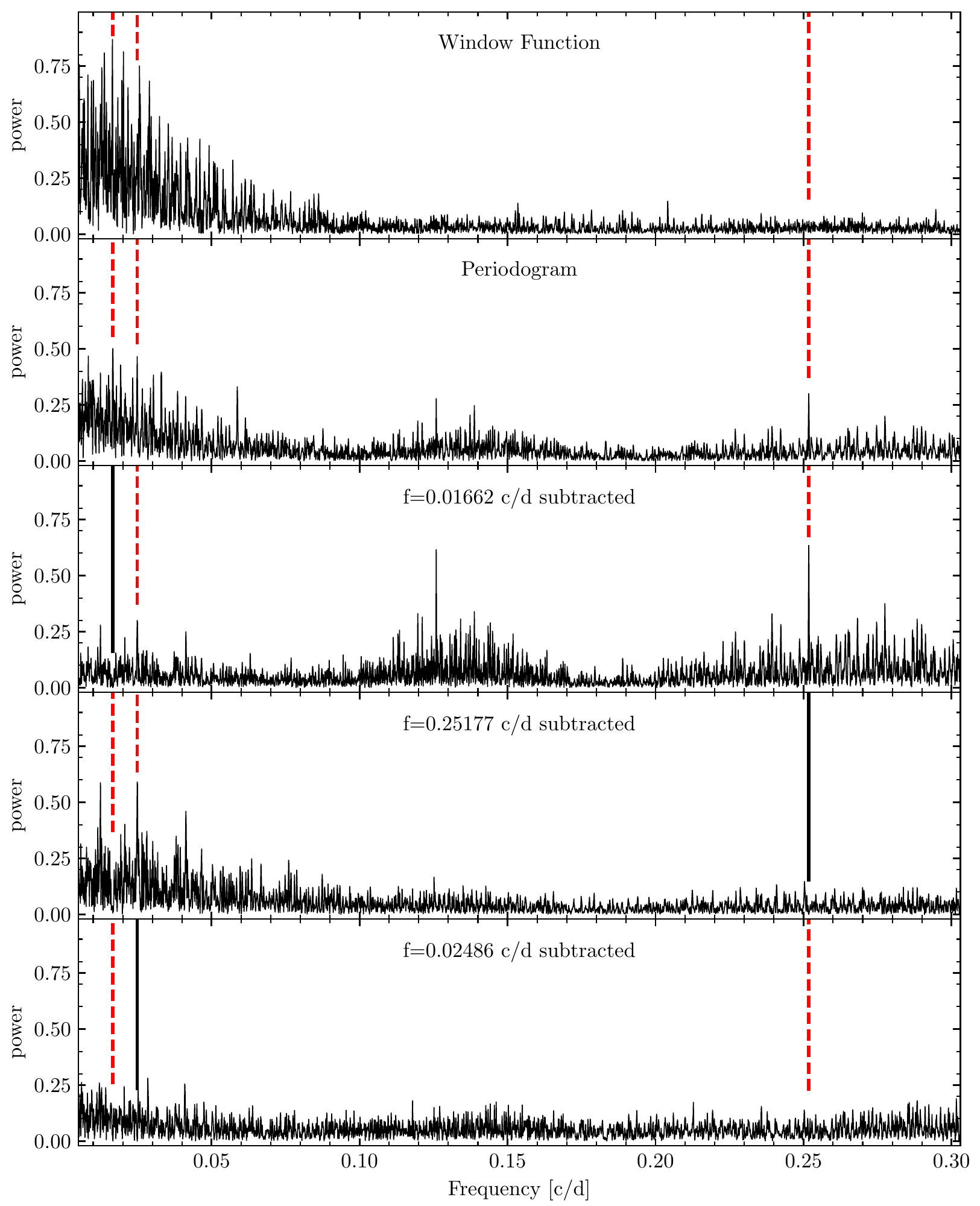}
\caption{Window function (top) and Lomb-Scargle periodograms of the successively pre-whitened {\it Hermes} BIS measurements. The three periods shown in Fig.\,\ref{fig:BISvariations} are indicated by vertical dashed red lines. The period subtracted in the previous step is marked by a solid black line in the following periodogram. Other notable peaks in the 2nd and 3rd panel from the top include the peak at twice \Ppuls\ (2nd panel) and at twice $P_{40d}$ (third panel). \label{fig:BISLombScargle}}
\end{figure*}

To more closely investigate additional periodicities in spectral data, it is desirable to consider quantities that provide precise information on a star's intrinsic variability without being sensitive to orbital motion. The bisector inverse span (BIS) provides a simple and precise quantification of spectral line profile asymmetry that varies on the timescale of \Ppuls\ in all of the hundreds of Cepheids observed by the Geneva Cepheid Radial Velocity Survey (Anderson, et al., in prep.). BIS is defined as the velocity difference between the top and bottom parts of a CCF, see \citep[henceforth: A16]{2016MNRAS.463.1707A} for a detailed discussion of BIS variability in the long-period Cepheid $\ell$~Carinae. BIS measurements  benefit from a much increased signal-to-noise ratio of CCFs compared to bisector measurements of individual spectral lines, and we estimate their short-term precision to be approximately $6-10$\ms\ based on groups of consecutive observations (cf. Appendix \ref{app:HermesRVs}). Hence the {\it Hermes} BIS measurements are a factor of $2-3$ more precise than the {\it Hermes} RV measurements. This precision gain results from the fact that each BIS measurement is a \textit{differential} quantity determined on a single observation. As a result, BIS is virtually unaffected by changes in the wavelength scale due to changing ambient pressure, which is the dominant RV uncertainty for {\it Hermes} data. Moreover, BIS is unaffected by orbital motion, which does not contribute to line asymmetry. In the case of the extremely high S/N CCFs of Polaris, the main systematic effect on BIS is the stability of the instrumental line profile, which is much better than the long-term wavelength-scale stability. Hence, BIS measurements are extremely well-suited to investigate additional periodicities intrinsic to Polaris.

We employ Lomb-Scargle periodograms \citep{1976Ap&SS..39..447L,1982ApJ...263..835S} computed using the \texttt{astropy} package \citep[Version 3.1]{astropy:2018} to search for additional periodicities in BIS data. We first compute periodograms for frequencies corresponding to periods between one year and half a day. After an initial inspection, we focus our attention to frequencies shorter than $0.3$\,cycles/d. Inspired by the second-order Fourier series model that best fits the RV data and the typically non-sinusoidal BIS variations in Cepheids (e.g. A16), we compute the periodograms using two-term Fourier series. We carry out a standard pre-whitening procedure, successively searching for  dominant peaks in the periodogram and removing a combined model of all previously identified frequencies before computing new residual periodograms. The modeled frequencies are determined as the frequency at the highest peak. We adopt the following functional form for  this multi-periodic BIS model: 
$v_{\rm{bis}} (t) = v_{0,\rm{bis}} + \sum_{i,j} [ a_{i,j}\cdot\cos(2\pi j (t-E)/P_i) + b_{i,j}\cdot\sin(2\pi j (t-E)/P_i)]$, 
where $v_{\rm{bis}}$ denotes BIS velocities, $a$ and $b$ the coefficients of each Fourier harmonic $j$, $P_i$ the periods of each signal, $t$ observation date, and $E$ the reference epoch. 
Figure\,\ref{fig:BISLombScargle} illustrates our results, starting with the window function of the BIS measurements, the first periodogram and three residual periodograms.

The highest peak in the BIS periodogram occurs at 0.01662 c/d, i.e., at a period of $60.16582$\,d. After removing this signal, we find a dominant peak at the well-known pulsation period, \Ppuls, together with its 0.5 c/d alias. After subtracting the combined model of the $60$\,d and \Ppuls\ signals, we find two nearly equally high peaks at $0.02486$\,c/d and its $0.5$\,d alias (identified by phase folding the data with either signal). The successive removal of the three frequencies near $60.17$\,d, \Ppuls, and $40.22$\,d reduces the RMS of the BIS measurements from $223$\,\ms\ to $135$\,\ms\ and $67$\,\ms, respectively. The remaining RMS is relatively high compared to the precision estimate of the BIS measurements, leaving the possibility of there being additional signals beyond this three-period model. However, given the uncertain nature of the dominant $60$\,d signal, we conclude that additional observations are required to further investigate the multi-periodicity of Polaris. To test the significance of our results, we re-ran the analysis using sub-sets of the full data set, such as all even and all odd numbered observations, and confirmed the presence of the identified frequencies. We also repeated the analysis after splitting the dataset in half (in time). The second half clearly reveals these signals, thanks to the dense sampling over a relatively short time, cf. Fig.\,\ref{fig:RVcurve}, whereas the results based on the first half of the data set are less clear.

\begin{figure*}
\centering
\includegraphics{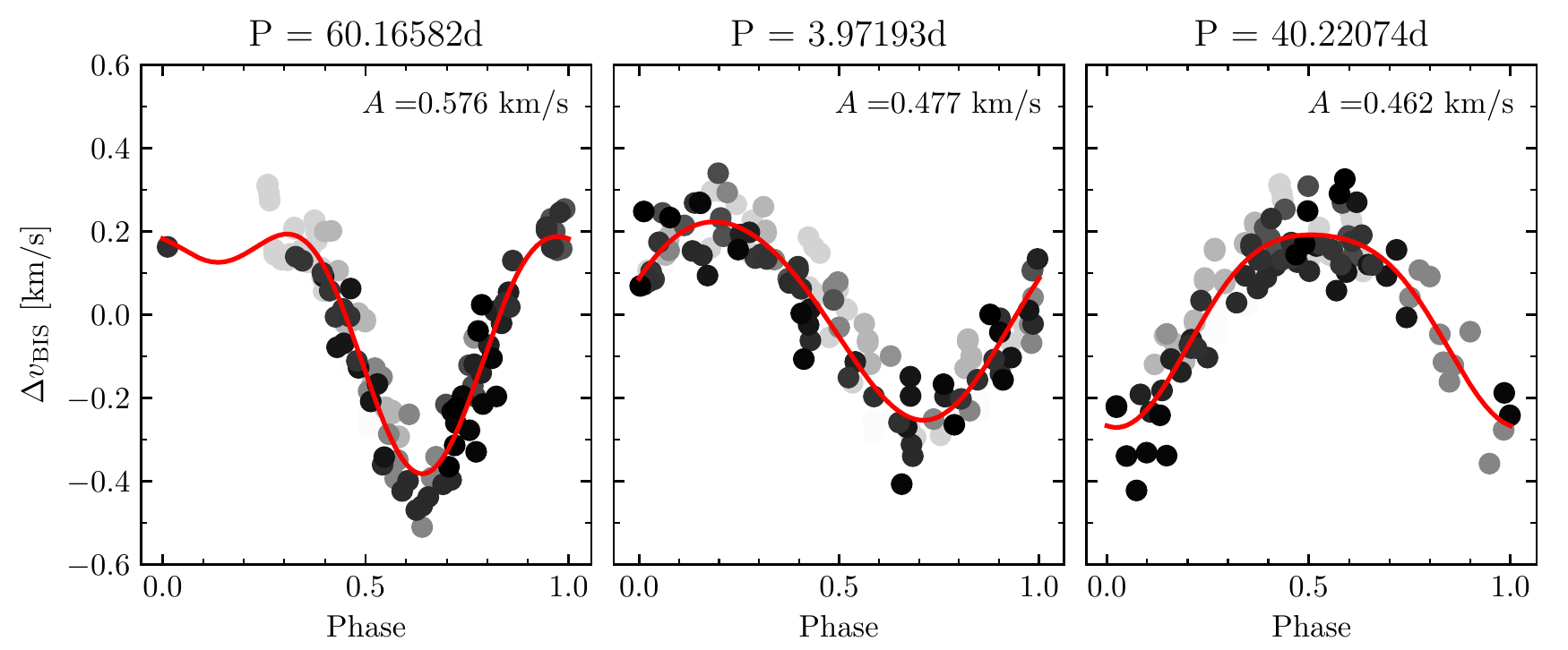}
\caption{BIS variations of Polaris measured from {\it Hermes} spectra phase-folded using three periods determined from Lomb-Scargle periodograms. Each panel shows the BIS variations associated with the periods indicated at the top ($\Delta v_{\rm{BIS}}$). The signals at $P_{60d} = 60.16582$d, \Ppuls$=3.97139$d, and $P_{40d} = 40.22074$d were identified using a successive pre-whitening procedure in this order. Second-order Fourier series models were fitted to the data using all three periods simultaneously and are shown separately for each signal as a solid red line. $P_{60d}$ and $P_{40d}$ are unlikely to be independent frequencies, since $P_{40d}/P_{60d} = 0.6685$ is suspiciously close to $2/3$ while there remain phase gaps in the sampling of the $60$\,d signal. The signal at $40.2$\,d had previously been identified in bisector variations by HC00. We notice that the $40$\,d and $4$\,d signals have similar amplitude. The grayscale color coding traces observing date (white to black).\label{fig:BISvariations}}
\end{figure*}

Figure\,\ref{fig:BISvariations} shows the BIS data phase-folded with each of these three periods after subtracting the model corresponding to the other two frequencies. As can be seen by the peak-to-peak amplitudes provided in each panel, the $60$\,d signal is the strongest of the three as expected from it having the highest peak in the BIS periodogram. However, it is suspicious that the ratio of the two periods $40.22074/60.16582 = 0.66850 \approx 2/3$, while the $60$\,d signal is not fully phase-sampled, cf. Fig.\,\ref{fig:BISvariations}. Conversely, the signal at \Ppuls\ and at $40.2$\,d are densely phase-sampled. The reality of the $40.2$\,d signal is strongly supported by HC00's prior discovery based on an independent data set obtained in the years 1992-1993 with a different window function. We therefore consider the $40.22$\,d frequency a true signal of Polaris, whereas we caution that confirmation of the $60$\,d signal is needed. It is remarkable that the amplitude of the $40.22$\,d signal is essentially equal to the BIS signal at \Ppuls\ even though its RVs counterpart is much weaker. Therefore, the $40.22$\,d signal primarily affects line shape, not line position, providing important constraints on the physical mechanism behind this signal, which may contribute to a clearer understanding of Polaris' puzzling properties.

\subsection{Is Polaris a Fast Rotator?\label{sec:disc:rotation}}

HC00 considered three different possible of origins of the $40$\,d bisector signal: macroturbulent spots, (magnetic) cool spots, and non-radial pulsations. Although they found no detailed model that could reproduce all the observed features, HC00 considered non-radial pulsations the most likely explanation, noting multiple additional periodicities reported in the literature as the most compelling reason for doing so. HC00 further explained that their model invoking surface inhomogeneities would imply high levels of magnetic activity, for which no evidence was available at the time. In particular, Polaris had not yet been detected in X-rays. 

Meanwhile, X-rays from Polaris have been detected using {\it Chandra} \citep{2010AJ....139.1968E}, with $L_X = 28.89\ \rm{erg s}^{-1} (0.3 -8 \rm{keV})$, which was recently updated to $L_X = 28.8 - 29.2\ \rm{erg s}^{-1} (0.1-0.6 \rm{keV})$ \citep{2017ApJ...838...67E}. Additionally, detailed predictions of Cepheid properties based on Geneva stellar evolution models that incorporate the effects of rotation \citep{2014A&A...564A.100A,2016A&A...591A...8A} can now be compared to Polaris' observed properties, and in particular to investigate whether the $40.2$\,d BIS signal could originate in rotation.

To investigate the rotation hypothesis, we assume that the equatorial rotation period $P_{\rm{rot}} = 40.2$d. The interferometrically measured diameter $3.123$\,mas \citep{2006A&A...453..155M} projected to the distance implied by the {\it Gaia} DR2 parallax of Polaris B of $7.292$\,mas implies an equatorial rotation velocity $v_{\rm{rot}} = 58.0$\,\kms. Using $v_{\rm{rot}}\sin{i} = 8.4$\,\kms\ from HC00 yields a very low inclination of $i \sim 8.3^\circ$. Thus, if the $40.2$\,d BIS signal were due to rotation, then Polaris would be seen nearly pole-on.

However, predictions based on Geneva models demonstrate that the rotation hypothesis is not consistent with observed properties of Polaris. The closest match to the implied surface rotation velocity is achieved by a $5 M_\odot$ first overtone Cepheid on the first instability strip crossing with Solar metallicity and very fast initial rotation ($\Omega / \Omega_{\rm{crit}} = 0.9$) near the hot edge of the instability strip, with $v_{\rm{rot}} = 55.6$\kms\ \citep[their Tab.\,A4]{2016A&A...591A...8A}. Observed CNO abundance ratios of $[N/C] = 0.59$ and $[N/O] = 0.42$ \citep{2005MNRAS.362.1219U} also agree with  predictions for this model, cf. Figs. 10 and 11 in \citet{2014A&A...564A.100A}.
However, neither the radius nor the absolute V-band magnitude of such a model are consistent with observations, with $26\,R_\odot$ predicted vs. $46\,R_\odot$ from interferometry and $V = -2.6$\,mag predicted vs. $-3.73$\,mag observed \citep{2018ApJ...863..187E}. To reach larger radii and greater luminosities would imply larger mass, which is disfavored by the recent astrometric mass measurement \citep[$3.45 \pm 0.75\,M_\odot$]{2018ApJ...863..187E}. Moreover, higher mass would tend to reduce the surface rotation velocity during the Cepheid evolutionary phase.

We therefore conclude that a) rotation does not cause the $40.2$\,d BIS signal, and b) that the latest empirical mass and radius estimates, which assume the highly precise {\it Gaia} DR2 parallax of Polaris B, cannot be explained by ordinary single star evolution models. This is intriguing, since observed period-radius relations \citep{2018ApJ...862...43P} and several other observables, incl. period-luminosity relations \citep{2016A&A...591A...8A}, agree closely with predictions based on the Geneva models. Alternative explanations of the intriguing $40.2$\,d BIS signal include non-radial pulsations as well as interactions between  pulsations and the convective envelope (cf. A16 for modulated variability due to convection-pulsation interaction). Additional monitoring and analysis of Polaris' BIS variability is required to further constrain the origin of this signal.

\section{Conclusions \label{sec:conclusions}}

We present 161 high-precision RVs of the Cepheid Polaris Aa, measured on high-resolution optical spectra collected between 2011 and 2018. We investigate the stability of the {\it Hermes} RV curve over this duration using an empirical RV template fitting technique and demonstrate a stable RV amplitude to within $\sim 30$\ms\ over the 7 year timespan of the observations. The precise {\it Hermes} data provide evidence for additional RV signals that reveal themselves as slow linear trends over the typical $10$\,d duration of each observing epoch. The peak-to-peak amplitudes associated with these signals are on the order of $\sim 100$\ms.

Applying the RV template fitting method to publicly accessible literature data, we find systematically different amplitudes between different data sets. However, no clear amplitude variations are recovered in any of the separately analyzed homogeneous data sets. We discuss the possibility of data inhomogeneity having contributed to reported RV amplitude differences in the literature. Additionally, we find strong indications that previously reported amplitude increases based on {\it SMEI} photometry were dominated by instrumental effects, which require detailed further investigation. Overall, we caution that Polaris' amplitude may be and may have been much more stable than previously thought.

We determine an updated solution to the $29.32$\,yr orbit of Polaris based on pulsation-averaged velocities measured via the template fitting method applied to {\it Hermes} data and one large homogeneous literature data set (K96-08). The derived systemic velocity of the Polaris Aa-Ab system agrees  with the {\it Gaia} DR2 measurement of the visual companion Polaris B to within the uncertainties.

We confirm line bisector variations with a period of $40.22$\,d that were originally discovered by HC00 using an independent set of bisector measurements of individual spectral lines. An additional periodicity of $60.17$\,d is indicated by the periodogram analysis, although we caution that this period may not be an independent frequency due to incomplete phase sampling and a suspicious period ratio of nearly $2/3$ with the $40.22$\,d signal. We rule out a rotational origin of the $40.22$\,d BIS signal, which is most likely caused by non-radial pulsations or interactions between pulsations and the convective envelope.

Polaris is a truly enigmatic type-I Cepheid. By reconsidering and revising the puzzling accounts of its putative amplitude variations and additional periodicities, we have sought to  facilitate a clearer understanding of this important star's properties. Additional spectroscopic and photometric monitoring are now needed to fully unravel its multi-periodic signals, and new interferometric observations are required to better constrain its circumstellar environment. Both of these avenues should be pursued in order to explain Polaris' properties and achieve a clearer understanding of the evolutionary paradigm of classical Cepheid variable stars.

\begin{acknowledgements} 

This work would not have been possible without the help of several observers, including Lovro Palaversa, Berry Holl, Maria S\"uveges, Micha\l\ Pawlak, Andreas Postel, Kateryna Kravchenko, Maroussia Roelens, Nami Mowlavi, and May Gade Petersen. The competent and friendly assistance of the Mercator support staff and staff at KU Leuven's astronomy department (in particular Jesus Perez Padilla, Saskia Prins, Florian Merges, Hans van Winckel, and Gert Rasking) is much appreciated.\\

RIA thanks the anonymous referee for a timely and constructive report that helped to improve the manuscript. The author is furthermore pleased to extend a warm thanks to Dietrich Baade and Antoine M\'erand for interesting discussions, as well as Byeong-Cheol Lee and Valery Kovtyukh for communications regarding Polaris. \\

This research is based on observations made with the Mercator Telescope, operated on the island of La Palma by the Flemish Community, at the Spanish Observatorio del Roque de los Muchachos of the Instituto de Astrof\'isica de Canarias. {\it Hermes} is supported by the Fund for Scientific Research of Flanders (FWO), Belgium, the Research Council of K.U. Leuven, Belgium, the Fonds National de la Recherche Scientifique (F.R.S.-FNRS), Belgium, the Royal Observatory of Belgium, the Observatoire de Gene\`eve, Switzerland, and the Th\"uringer Landessternwarte, Tautenburg, Germany. \\

This research has made use of NASA's Astrophysics Data System; the SIMBAD database and the VizieR catalog access tool\footnote{\url{http://cdsweb.u-strasbg.fr/}} provided by CDS, Strasbourg; Astropy, a community-developed core Python package for Astronomy \citep{2013A&A...558A..33A,astropy:2018};
\end{acknowledgements}

\bibliographystyle{aa} 
\bibliography{biblio}

\begin{appendix}

\section{{\it Hermes} RVs of Polaris Aa \label{app:HermesRVs}}
Table \ref{tab:app:RVdata} lists the {\it Hermes} measurements presented here. The full list of measurements is available in the online appendix of the journal and via the CDS.

\begin{table}[h]
\centering
\begin{tabular}{@{}lrrrr@{}}
\hline
BJD  - $2.4 \times 10^6$ &  $v_{\rm{r}}$  &  $\sigma_{v_{\rm r}}$ & $v_{\rm{BIS}}$ & {\it Hermes} ID \\
d & \kms & \kms & \kms \\
\hline
55816.544945 & -17.573 & 0.015 & -1.407 & 00373294 \\
55816.545666 & -17.592 & 0.015 & -1.414 & 00373295 \\
55817.602438 & -18.716 & 0.015 & -1.027 & 00373387 \\
55817.603159 & -18.728 & 0.015 & -1.017 & 00373388 \\
55817.603879 & -18.713 & 0.015 & -1.019 & 00373389 \\
\hline
\multicolumn{5}{c}{} \\
\multicolumn{5}{c}{Full table available at CDS} \\
\multicolumn{5}{c}{} \\
\hline
58438.454099 & -10.741 & 0.015 & -1.667 & 00902032 \\
58439.308096 & -12.491 & 0.015 & -1.303 & 00902615 \\
58439.603573 & -13.307 & 0.015 & -1.128 & 00902733 \\
58442.781257 & -11.286 & 0.015 & -1.233 & 00903141 \\
58443.315679 & -12.683 & 0.015 & -0.964 & 00903383 \\
\hline
\end{tabular}
\caption{{\it Hermes} RV and BIS measurements used in this work. The full table is available in the online appendix and at the CDS. A fixed precision of $15$\ms\ is adopted for RVs, whereas the instrumental zero-point is estimated to be stable at the level of $50-70$\ms\ over the duration of the measurements, cf. \S\ref{sec:obs}. BIS measurements have a typical short-term precision of $\sigma_{\rm{BIS}} \sim 6-10$\ms\ as indicated by the standard deviation of BIS values determined using groups of consecutive measurements. \label{tab:app:RVdata}}
\end{table}

\section{Figures Showing RV Template Fitting Method Applied to Literature RVs\label{app:Kamper}}

\begin{figure*}
\centering
\includegraphics{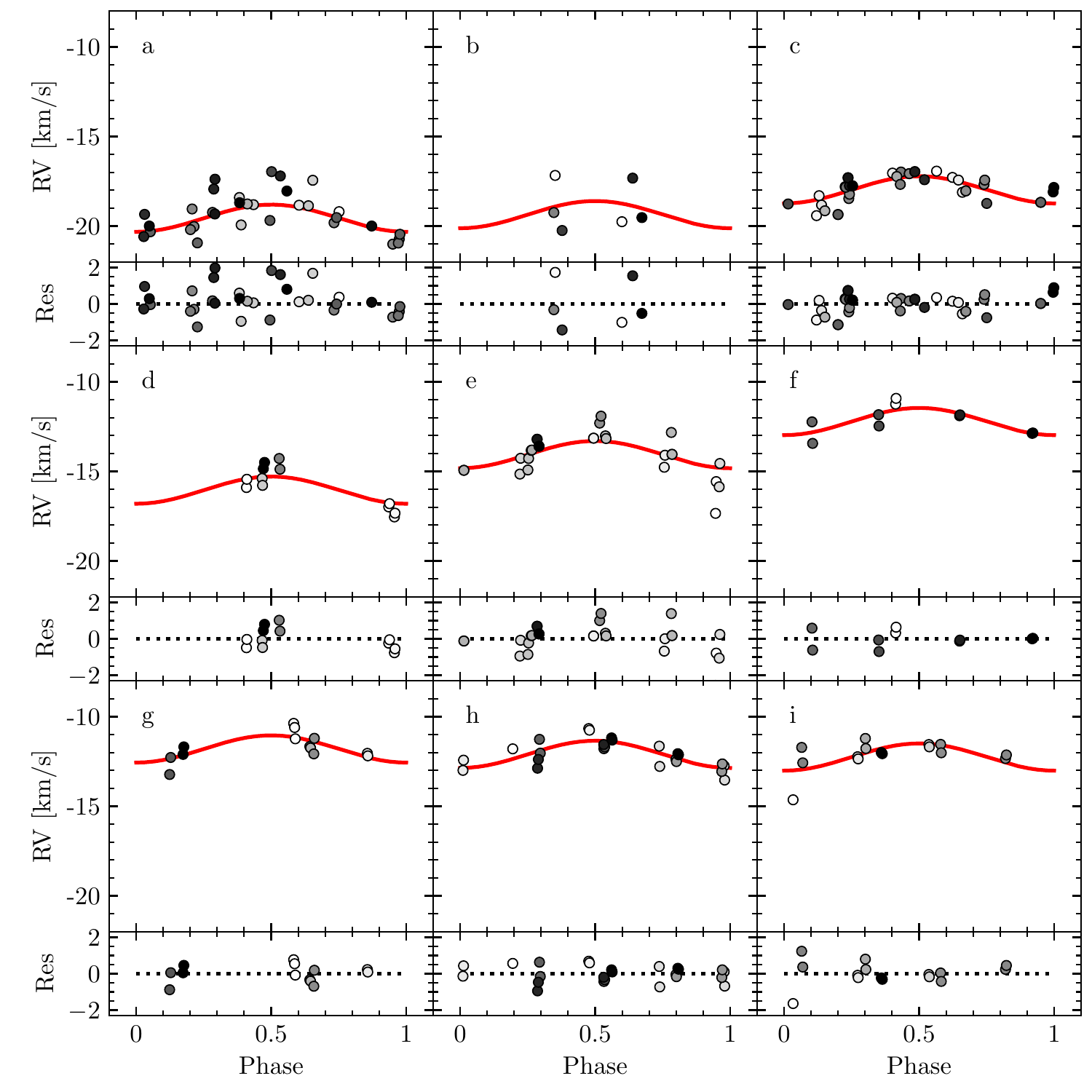}
\caption{RV template fitting method applied to high-dispersion photographic RVs by \citet{1996JRASC..90..140K} (K96-08). No significant amplitude variation is seen with time, although the orbital motion is clearly apparent.\label{app:fig:Kampertemplatefit}}
\end{figure*}

\begin{figure*}
\centering
\includegraphics{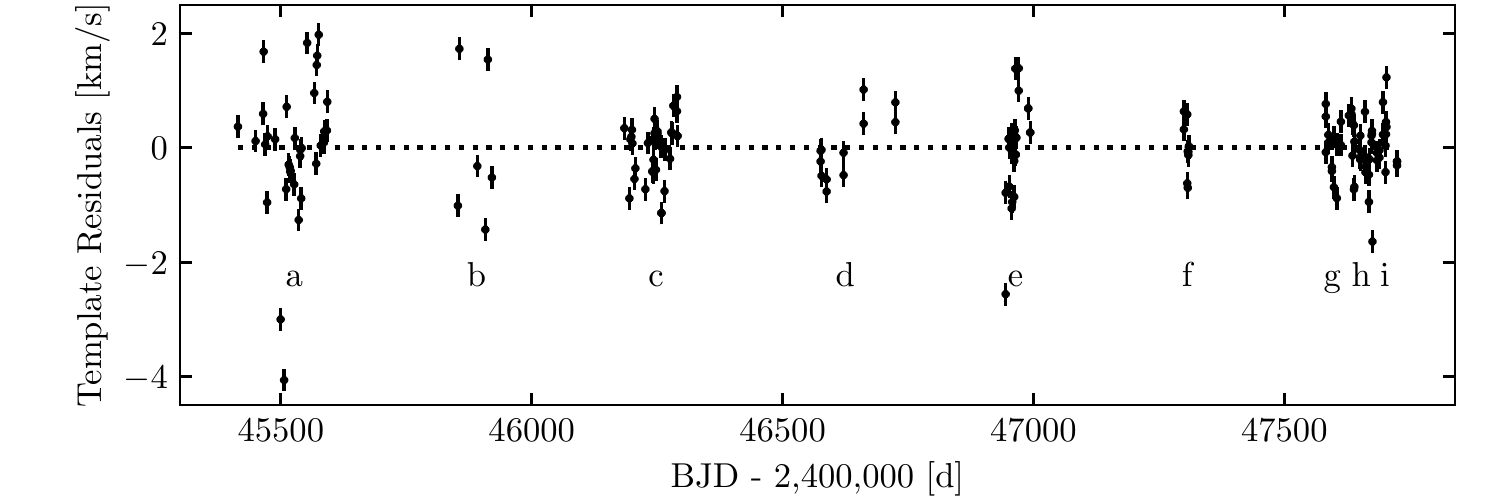}
\caption{Residuals from RV template fitting method applied to K96-08 RVs. The improvement of the RV precision following `standardization' of the RV measurement following \citet{1983ApJ...274..755A} is clearly seen. \label{app:fig:Kampertemplateresiduals}}
\end{figure*}

\begin{figure*}
\centering
\includegraphics[scale=0.9]{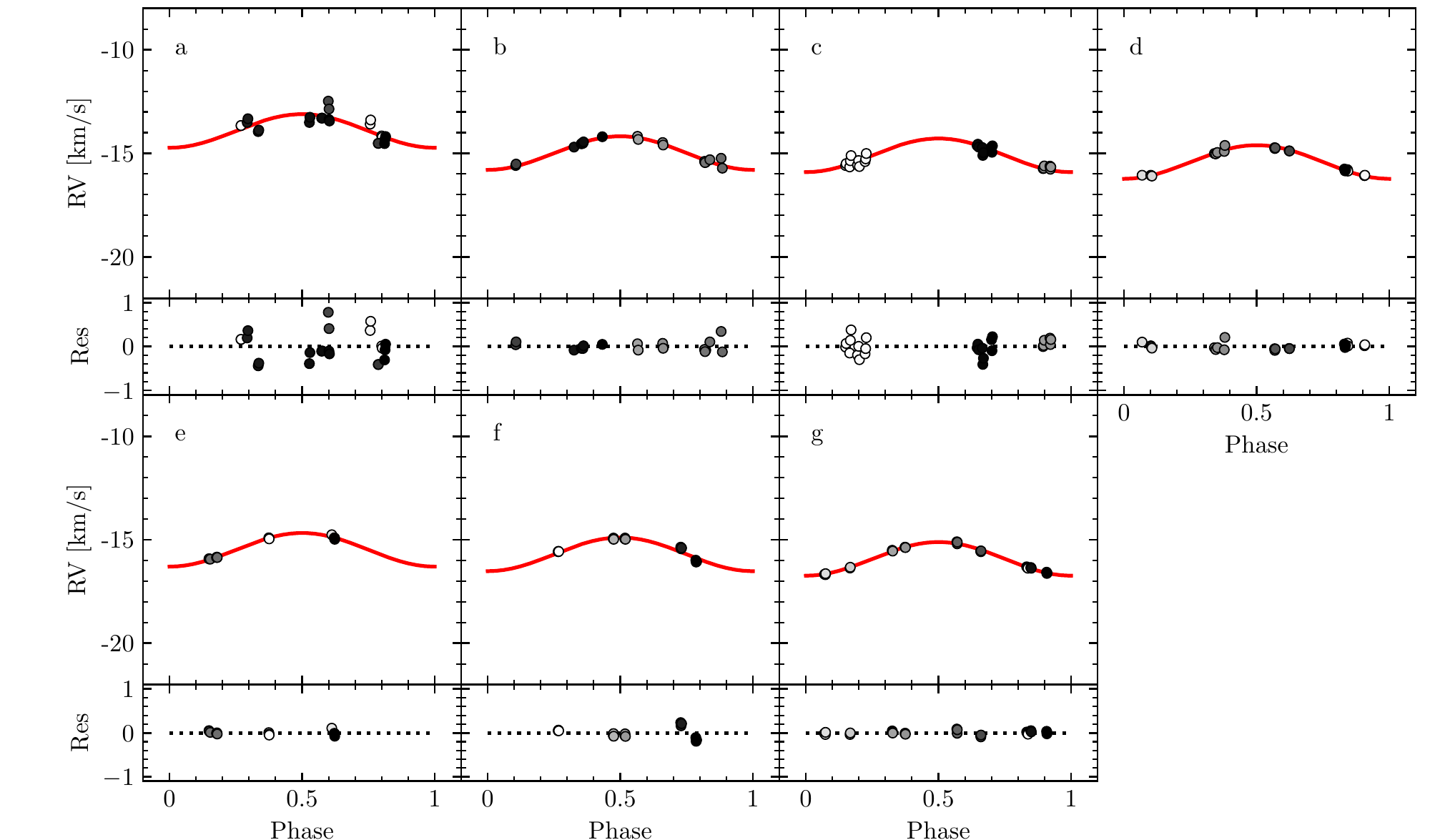}
\caption{RV template fitting method applied to K96-CE RVs. No significant variation in amplitude is apparent with time, although the scatter in the first three epochs is significantly larger than in the later four, likely because of calibration issues.\label{app:fig:K96CEtemplatefit}}
\end{figure*}

Figure \ref{app:fig:Kampertemplatefit} shows the RV template fits applied to K96-08 data; the residuals as a function of time are shown in Fig. \ref{app:fig:Kampertemplateresiduals}. Figure \ref{app:fig:K96CEtemplatefit} shows the template fits to the K96-CE data.

\section{{\it SMEI} Per-epoch Fourier Series Fits\label{app:SMEI}}
\begin{figure*}
\centering
\includegraphics{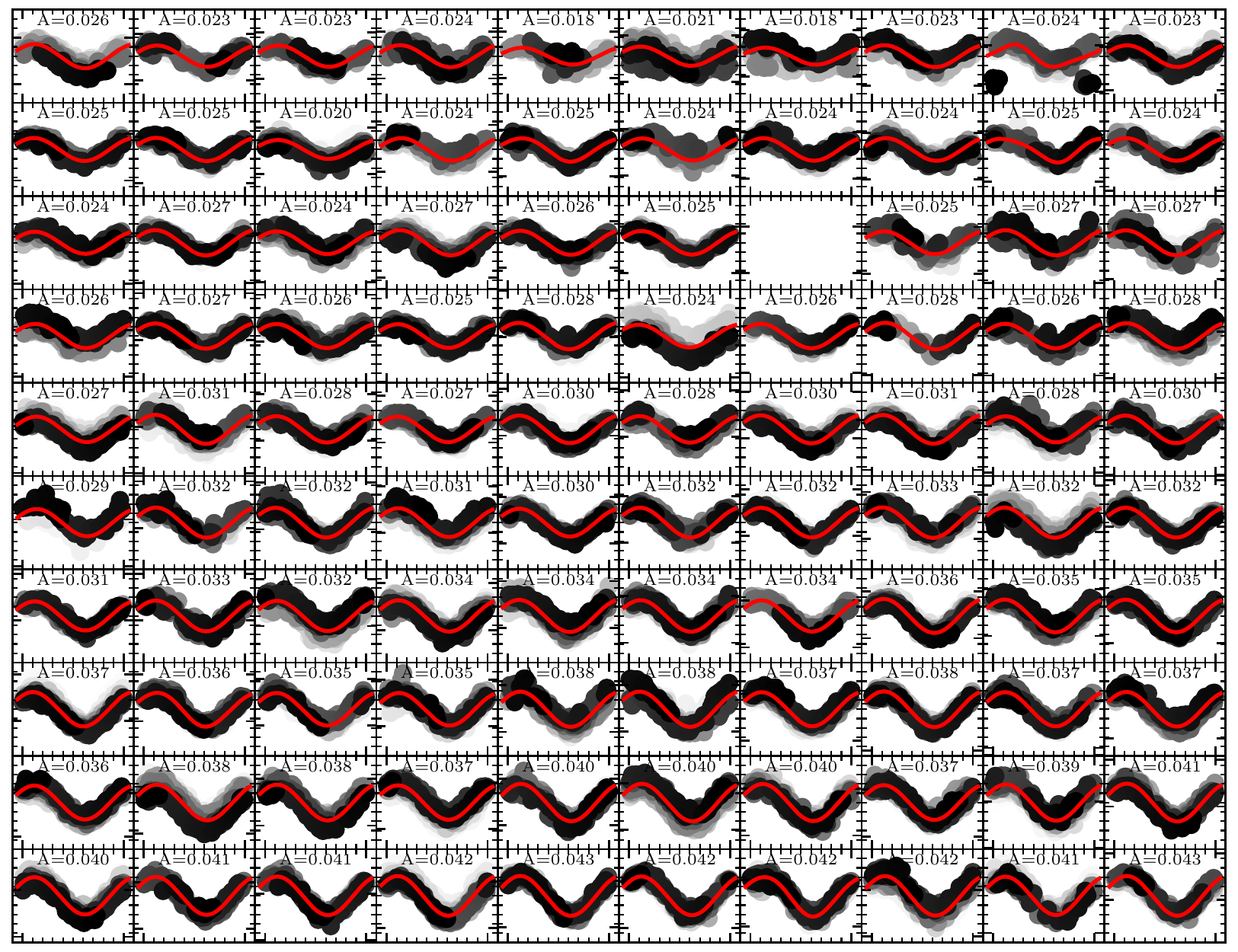}
\caption{Second-order Fourier series models fitted to 100 sub-epochs of {\it SMEI} data, cf. \S\ref{sec:disc:SMEI}. Insufficient data were available for the 27th epoch. Peak-to-peak amplitudes are indicated in each panel. Time increases across the figure from left to right and top to bottom. In each panel, observation dates are grayscaled from white to black. The ordinate ({\it SMEI} magnitude) and  abscissa (phase) ranges are $-0.05 - 0.05$\,mag and $-0.1 - 1.1$, respectively, in all panels.\label{fig:SMEIepochfits}}
\end{figure*}

Figure\,\ref{fig:SMEIepochfits} shows the Fourier series fits to the 100 sub-epochs of {\it SMEI} observations used in \S\ref{sec:disc:SMEI}. 

\end{appendix}

\end{document}